\documentclass[lettersize,journal]{IEEEtran}
\usepackage{soul}
\usepackage{graphicx}%
\usepackage{multirow}%
\usepackage{amsmath,amssymb,amsfonts}%
\usepackage{amsthm}%
\usepackage{mathrsfs}%
\usepackage{xcolor}%
\usepackage{textcomp}%
\usepackage{manyfoot}%
\usepackage{booktabs}%
\usepackage{tcolorbox}
\usepackage{algorithm}%
\usepackage{listings}%
\usepackage{hyperref}
\usepackage{version}
\usepackage{bm}
\usepackage{xspace}
\usepackage{comment}
\usepackage{float}
\usepackage{array,multirow}
\usepackage{etoolbox}
\usepackage{siunitx}
\usepackage{rotating}
\usepackage{subcaption}
\usepackage{xr-hyper}
\AtBeginDocument{
  \let\oldcite\cite
  \renewcommand{\cite}[1]{\texorpdfstring{\oldcite{#1}}{[citation]}}
}

\includeversion{transactionsReferences}
\excludeversion{natureReferences}

\includeversion{mainVersion}

\includeversion{appendixVersion}

\includeversion{allFigs}
\includeversion{allTabs}

\includeversion{nature}
\excludeversion{longNature}

\includeversion{alternativeText}
\excludeversion{alternativeText}

\excludeversion{alternativeFigs}
\usepackage{lineno}

\usepackage{multibib}
\newcites{appendix}{Appendix References}

\def\nitbf{\noindent\textbf}

\usepackage{verbatim}

\begin{document}

\newcommand{\fred}[1]{{\color{orange}{[[FG: #1]]}}}
\newcommand{\tiago}[1]{{\color{teal}{[[TB: #1]]}}}
\newcommand{\ramon}[1]{{\color{red}{[[RAP: #1]]}}}
\newcommand{\joanna}[1]{{\color{green}{[[JM: #1]]}}}
\newcommand{\davide}[1]{{\color{blue}{[[DF: #1]]}}}
\newcommand{\davidedone}[1]{{\color{cyan}{[[DF(done): #1]]}}}
\newcommand{\f}[1]{{\color{blue}{#1}}}
\newcommand{\fc}[1]{{\color{magenta}{#1}}}
\newcommand{\cut}[1]{{\color{red}{[[cut: #1]]}}}
\newcommand{\display}[1]{{\color{red}{#1}}}
\newcommand{\finalcut}[1]{}

\tcbset{colframe=black, colback=lightgray, arc=2mm}

\title{Small is Sufficient: Reducing the World AI Energy Consumption Through Model Selection}
\author{Anonymous}
\author{\IEEEauthorblockN{Tiago da Silva Barros, Frédéric Giroire, Ramon Aparicio-Pardo, and Joanna Moulierac}
\thanks{
This work has been supported by the French government National Research Agency (ANR) through the UCA JEDI (ANR-15-IDEX-01) and EUR DS4H (ANR-17-EURE-004), by the France 2030 program under grant agreements No. (ANR-23-PECL-0003 and ANR-22-PEFT-0002), through the 3IA Cote d’Azur Investments in the project with the reference number ANR-23-IACL-0001, by SmartNet, and by the European Network of Excellence dAIEDGE under Grant Agreement Nr. 101120726. (Corresponding authors: Tiago da Silva Barros, Frederic Giroire)

Tiago da Silva Barros is with Université Côte d'Azur (e-mail: Tiago.DA-SILVA-BARROS@univ-cotedazur.fr)

Frederic Giroire is with CNRS (e-mail: frederic.giroire@cnrs.fr)

Ramon Aparicio-Pardo is with Université Côte d'Azur (e-mail: Ramon.APARICIO-PARDO@univ-cotedazur.fr)

Joanna Moulierac is with Université Côte d'Azur (e-mail: Joanna.MOULIERAC@univ-cotedazur.fr)
}}









\def\setTasks{T}
\def\popularity{p}
\def\setModels{M}
\def\setModelsLarge{M^l}
\def\energy{E}
\def\utility{U}
\def\opt{o}

\def\ebefore{ E^{\text{before}}}
\def\eafter{ E^{\text{after}}}
\def\ubefore{ U^{\text{before}}}
\def\uafter{ U^{\text{after}}}

\def\energyReduction{\text{ER}}
\def\utilityDrop{\text{UD}}

\def\scenarioA{\textit{from largest to optimized}}
\def\scenarioB{\textit{all models considered}}

\def\best{{\tt best-performing}\xspace}
\def\efficient{{\tt energy-efficient}\xspace}
\def\bestShort{{\tt BE}\xspace}
\def\efficientShort{{\tt EE}\xspace}

\newcommand{\embf}[1]{\textbf{\textit{#1}}}

\maketitle
    
\begin{mainVersion}



\begin{abstract}
    The energy consumption and carbon footprint of Artificial Intelligence (AI) have become critical concerns due to rising costs and environmental impacts. In response, a new trend in green AI is emerging, shifting from the “bigger is better” paradigm, which prioritizes large models, to “small is sufficient,” emphasizing energy sobriety through smaller, more efficient models.

We explore how the AI community can adopt energy sobriety today by focusing on model selection during inference. Model selection consists of choosing the most appropriate model for a given task, a simple and readily applicable method, unlike approaches requiring new hardware or architectures. Our hypothesis is that, as in many industrial activities, marginal utility gains decrease with increasing model size. Thus,  applying model selection can significantly reduce energy consumption while maintaining good utility for AI inference. 

We conduct a systematic study of AI tasks, analyzing their popularity, model size, and efficiency. We examine how the maturity of different tasks and model adoption patterns impact the achievable energy savings, ranging from 1\% to 98\% for different tasks. Our estimates indicate that applying model selection could reduce AI energy consumption by 27.8\%, saving \SI{31.9}{TWh} worldwide in 2025 —equivalent to the annual output of five nuclear power reactors.
\end{abstract}



\section{Introduction}

The use of Artificial Intelligence (AI) has skyrocketed in recent years, transforming diverse fields as medicine, education, finance, robotics, games, etc., with amazing successes such as automatic writing, translation, driving, disease diagnosis, etc. 

These achievements have been made possible by the development of Deep Learning (DL) techniques, which leverages large-scale neural networks trained on vast amounts of data. 
The scale of machine learning models has expanded exponentially, from hundreds of parameters in the early 2000s to tens of millions in the 2010s, billions in the early 2020s, and now trillions in state-of-the-art large language models (LLMs) (see Figure~\ref{fig:epochAI:params_over_date}).

\begin{figure}[!h]
    \centering
    \includegraphics[width=\linewidth]{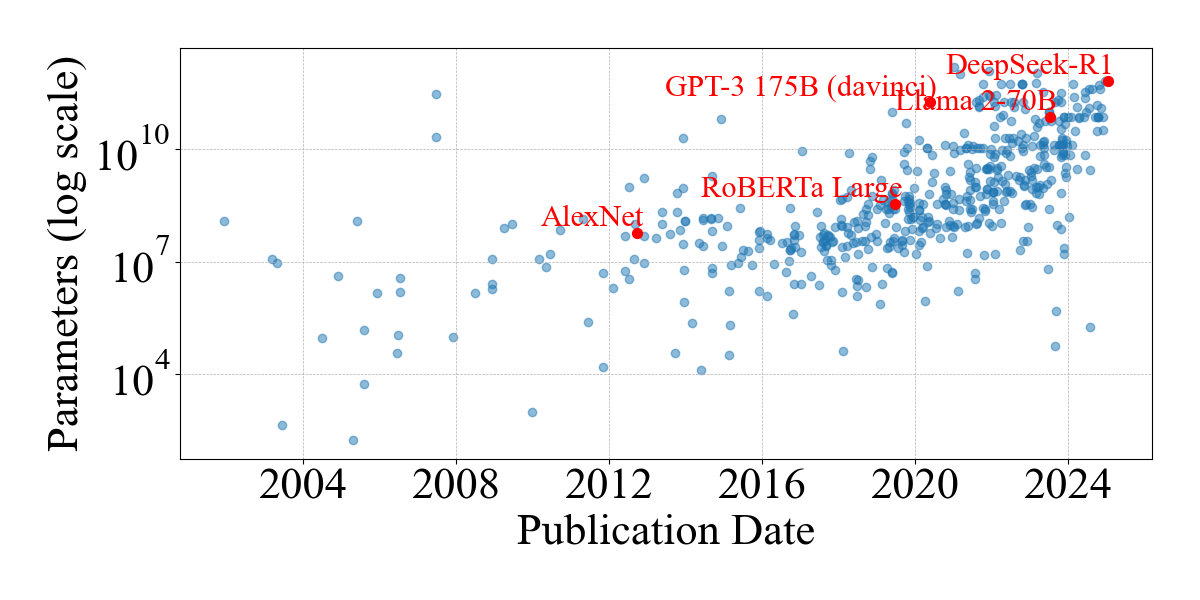}
    \caption{\textbf{Scaling Trends in AI Models.} Number of parameters (log scale) plotted against publication date for notable AI models, illustrating an exponential growth in model size over time. Data was extracted from \cite{EpochNotableModels2024}.}
    \label{fig:epochAI:params_over_date}
\end{figure}


The use of such models, coupled with the growth in data volumes and the infrastructure required to run them, with ever more powerful machines, has led to a rise in energy consumption \cite{wu2022sustainable}.
The US energy report \cite{USEnergyDataCenter} estimates that datacenters in the United States  will consume between 325 and 580 TWh in 2028, representing up to 12.0\% of the country total energy consumption,
where AI-related operations account for approximately 22\% of the datacenter consumption.

The rising energy demand of AI, coupled with the associated greenhouse gas emissions (GHGE), concerns different actors, including tech companies, governments, scientific community and the society; and poses a risk to the efforts for mitigating the climate changes impacts, such as established in the Paris Agreement \cite{paris2015paris} and in the Intergovernmental Panel on Climate Change (IPCC) report \cite{lee2023ipcc}.

Addressing AI energy consumption is, therefore, critical. 
Then, different methods have been proposed to reduce the energy consumption of AI targeting on the two main phases of a model life cycle: (i) training (e.g., hardware optimization \cite{tao2020challenges}, model compression \cite{zhang2018exploring}, architecture designing \cite{mehta2019espnetv2}),  and (ii) inference (e.g., model selection \cite{asperti2021survey,yu2022energy}). 
We focus on \textit{inference} here. Although the energy expended per inference task is orders of magnitude less than that of training, the sheer volume of inference requests results in a substantial cumulative impact. Indeed, inference has been estimated to account for approximately 60\% of ML energy usage~\cite{patterson2022carbon}.

In this work, we aim to investigate the \textit{impact of model selection} on inference AI energy consumption. Indeed, the release of new models such as DeepSeek LLM~\cite{deepseek} attracted attention to two potential paradigms for the future of the AI industry. The \textit{``bigger-is-better"}, in which everybody buys the best hardware (e.g., from NVIDIA) to run the best-performing (and usually the largest) models, and the \textit{``small-is-sufficient"}, in which users prioritize much smaller models with similar performance than the state-of-the art models. In this paper, we estimate the energy consumption of these two potential futures.


Energy-efficient model selection consists in selecting a model that is smaller than the state-of-the-art model to solve an AI task, but as close as possible in performance. Indeed, a large number of models exist to solve any AI task, and widely available platforms such as Hugging Face~\cite{huggingface} and Papers with Code~\cite{paperswithcode} facilitate access to diverse AI models, allowing practical implementation of model selection strategies. Model selection is thus a very simple method that can be immediately applied by any user who needs to solve an AI task. There is no need to retrain models, use specific hardware, implement complex architecture, etc.

\begin{figure}[!ht]
\centering
\includegraphics[width=\linewidth]{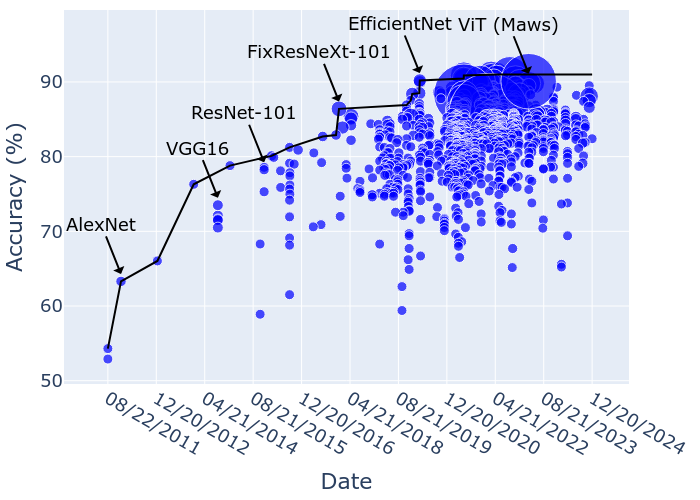}\hfill
\caption{\textbf{Evolution of \textit{Image Classification} models over time.} Accuracy (\%) of image classification models on the \textit{ImageNet} dataset plotted against their publication date. Each point corresponds to one model. Each data point represents a model, with the marker size proportional to the number of parameters. The plot shows an accuracy "plateau" post-2020, with the continued release of models across various sizes. Data extracted from the \textit{Papers With Code} platform.
\label{fig:evolutionImageClassImageNET}}
\end{figure}

If we take as an example the classic task of Image Classification, several thousand models have been introduced to solve it (\num{5134} models are available in Hugging Face and \num{4405} models in Papers with Codes), each 
having a different trade-off between its 
performance, its size, and its energy consumption. 
Figure \ref{fig:evolutionImageClassImageNET} illustrates the evolution of larger and more efficient models over time, from the first convolutional neural network, AlexNet~\cite{krizhevsky2012imagenet}, with \SI{65}{M} parameters, to image transformers with \SI{1}{B} parameters. 
We observe two main phases over time: a rapid increase in performance followed by a plateau from 2021 onwards. 
This is related to a general phenomenon in every economic activity, known as the law of diminishing returns, also referred to as the law of diminishing marginal productivity. This concept can be traced back to the 18th century, in the work of Jacques Turgot~\cite{turgot}. 
As the task matures, researchers have focused in developing efficient models keeping same performance, while reducing model size, using different methods such as model sparsification \cite{frantar2023sparsegpt}, quantization \cite{shao2023omniquant}, compression \cite{ma2023llm}. Thus, the impact of model selection for an AI task depends on its stage of development.

Our contributions are as follows. We investigate the energy savings that can be achieved by applying AI model selection globally.
To do this, we first identify the most commonly used AI tasks in data centers. For each task, we analyze benchmarks from the \textit{Hugging Face}~\cite{paperswithcode} and \textit{Papers with Code}~\cite{huggingface} platforms to assess model utility and size tradeoffs. 
By evaluating task maturity and model adoption, we identify key opportunities for energy-efficient AI. 
Then, we propose a methodology for estimating the energy consumption of AI models during inference, and
finally, we estimate the energy savings by applying model selection and redirecting AI inference requests to energy-efficient models. 

Our results show the \textit{huge impact of the task stage of development and the patterns of model adoption} on the potential energy reduction, with a range from 1\% to 98\% across the different AI tasks. Our projection indicates that \textit{applying model selection globally} could lead to a 27.8\% reduction in AI energy consumption  \textit{without much impact on the model utility.} On the contrary, \textit{always using the state-of-the-art models} to solve all tasks would \textit{increase the global AI energy consumption by 111\%}.

\section{Analysis of AI Tasks and Models}
\label{sec:potentialForEnergySobriety}

This section analyzes the potential for energy sobriety in AI inference. We begin by comprehensively investigating the most prevalent AI tasks in data centers, the models employed for these tasks, their adoption rates, and relevant benchmarks for comparison. Subsequently, for each identified task, we examine the relationship between model size, user adoption, and utility, aiming to pinpoint opportunities for reducing energy consumption through informed model selection.



\subsection{Identifying the More Popular AI Tasks}
\label{sec:AIbenchmarks:popularity}

With the rise of cloud computing and the evolution of computational resources, major cloud providers, such as Amazon, Google, and Microsoft now offer platforms for deploying AI models at scale. These platforms facilitate the widespread use of AI across various domains, making it essential to understand which AI tasks are most frequently deployed. 

To address this, we conducted an investigation to:
(i) identify the most commonly deployed AI tasks in data centers, and
(ii) analyze reliable benchmarks used to assess and compare AI models for these tasks.

Several studies have investigated the key AI tasks addressed by industry and academia. For instance, the 2024 AI Index Report \cite{perrault2024artificial} highlights trending AI tasks and benchmarks associated with notable models, reported in 
\begin{natureReferences}
    Supplementary Information Table~\ref{tab:AIIndexFields}.
\end{natureReferences}
\begin{transactionsReferences}
     Appendix E, available in the online supplemental material.
\end{transactionsReferences}Main listed tasks are: \textit{Language, Coding, Image Computer Vision, Video Computer vision, Reasoning, Audio, Agents, Robotics,} and \textit{Reinforcement Learning}. Similarly, Bommasani et al. \cite{bommasani2021opportunities} examined models trained on vast datasets and identified key areas of application, including \textit{language}, \textit{vision}, \textit{robotics}, \textit{reasoning (logic)}, and \textit{interaction}.


In parallel, some platforms, such as \textit{Papers With Code} and  \textit{Hugging Face}, propose to serve as collection of multiple models for several AI tasks.
The \textit{Papers With Code} platform provides a collection of scientific papers with available implementation across several topics in computer science, including Artificial Intelligence. The number of papers for each AI field is available in 
\begin{natureReferences}
    Supplementary Information
\end{natureReferences} 
Figure~\ref{fig:PWCNbOfPapersPerArea}.

\begin{allFigs}
\begin{figure}
    \centering
    \includegraphics[width=\linewidth]{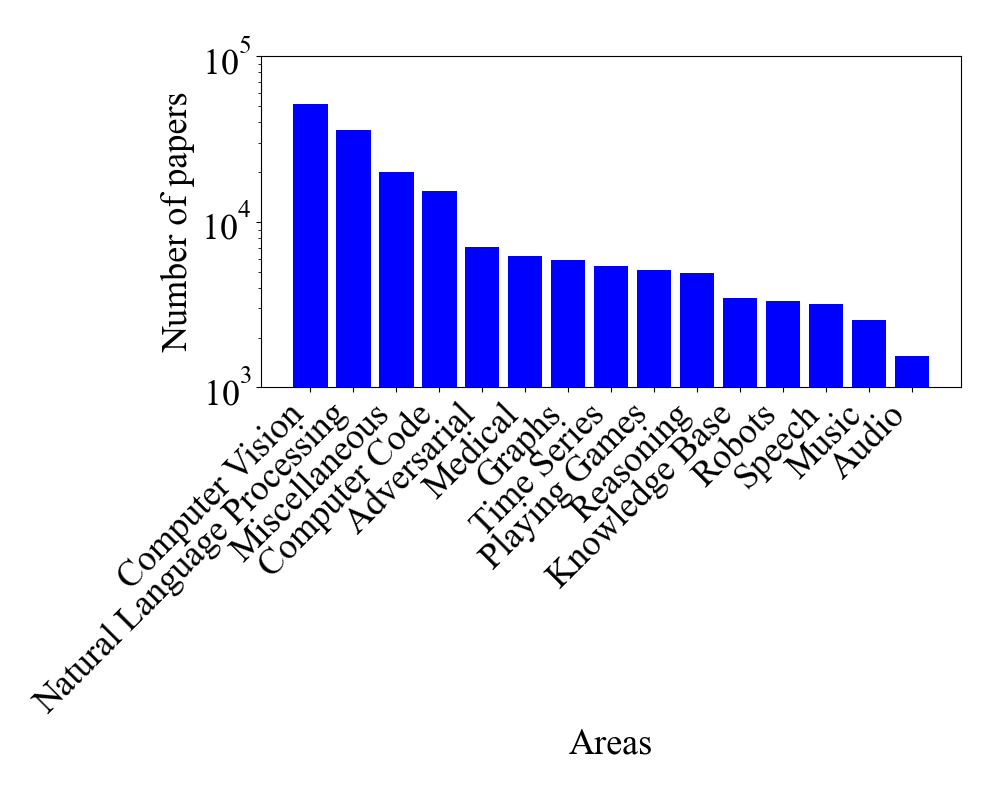}
    \caption{\textbf{Research Focus by Area on Papers With Code:} Numbers of papers across different AI research areas on the \textit{Papers With Code platform}. The results highlight significant interest in Computer Vision (image processing), Natural Language Processing, and a diverse range of Multimodal tasks categorized as Miscellaneous.
    \label{fig:PWCNbOfPapersPerArea}}
\end{figure}
\end{allFigs}
The \textit{Hugging Face} platform provides an open source library for hosting and deploying AI models with broad applicability. All AI tasks considered by \textit{Hugging Face} and their respective domains are described in 
\begin{natureReferences}
    Supplementary Information Table \ref{tab:listTasksFieldsHuggingFace}.
\end{natureReferences}
\begin{transactionsReferences}
     Appendix E, available in the online supplemental material.
\end{transactionsReferences}
. 

A \textbf{key metric for assessing the popularity of AI models} in \textit{Hugging Face} is the \textbf{number of downloads}, which reflects real-world adoption by users and developers.  Figure~\ref{fig:TaskPopulaity:HF_nbDownloads_field} describes the total number of downloads for the most popular AI fields and tasks. 
We observe that tasks belonging to \textit{language}, \textit{vision}, and \textit{audio} fields present a high level of adoption.

This observation is followed by the growing interest of industry and the academia in models for some high-impact tasks, e.g., text generation. 
Several models, such as \emph{ChatGPT} (OpenAI), \emph{Gemini} (Google), \emph{Deepseek}, and \emph{Llama} (Meta), have attracted a lot of interests by researchers and are widely deployed among several data centers. For instance, data from Meta point that the \emph{Llama} usage in major cloud service providers more than doubled from May through July 2024.

\begin{figure}[ht]
    \centering
    \includegraphics[width=\linewidth]{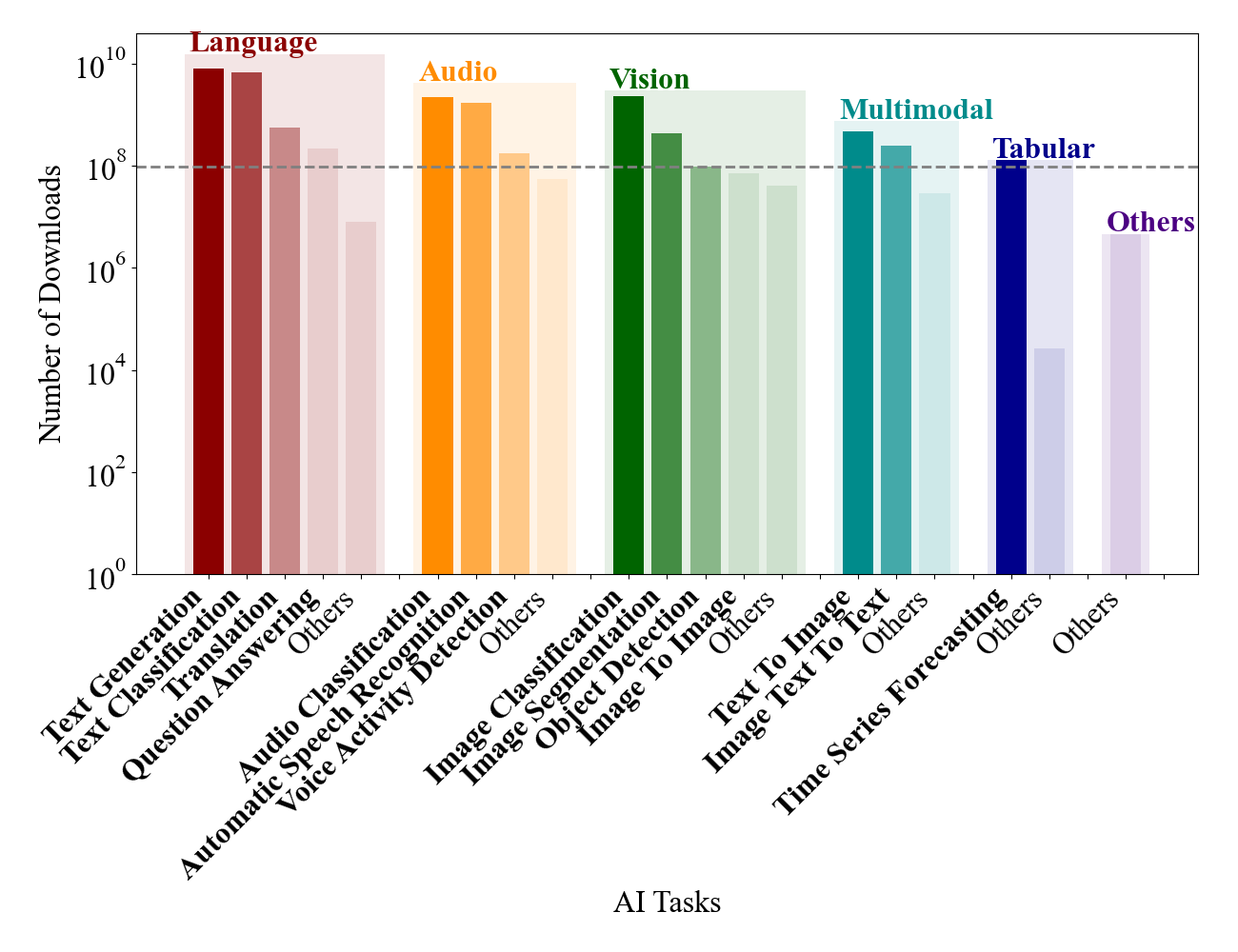}
    \caption{\textbf{Most popular AI fields and tasks.} Number of total model downloads on the \textit{Hugging Face} platform, categorized by AI field and AI task. The dashed gray line represents the download threshold used for task evaluation in this study, encompassing 97.3\% of all platform downloads.}
    \label{fig:TaskPopulaity:HF_nbDownloads_field}
\end{figure}


Based on the popularity analysis, we selected the most frequently used tasks in \textit{Hugging Face} platform to investigate the potential of AI energy sobriety. We focused on tasks with more than $10^8$  downloads in \textit{Hugging Face} platform.
The selected tasks grouped by their respective field are the following: 
\begin{itemize}
\item[] (i) for the language field: \textit{text generation, text classification, and translation}; 
\item[] (iii) for the audio field: \textit{automatic speech recognition, and audio classification}; 
\item[] (ii) for the vision field: \textit{image classification, object detection, image segmentation};
\item[] (iv) for the multimodal field: \textit{image-text to text and text to image}; and 
\item[] (v) for the tabular data field: \textit{time series forecasting}.
\end{itemize}

Moreover, we evaluate two additional text generation sub-tasks -\textit{mathematical reasoning} and \textit{code generation}- which are widely studied in literature and pointed as trending tasks in Artificial Intelligence. 

The tasks are described in
\begin{natureReferences}
    Methods \ref{Methods:benchmarksDescription}.
\end{natureReferences}
\begin{transactionsReferences}
Appendix E, available in the online supplemental material.
\end{transactionsReferences}

The above tasks present more than $22 \times 10^9$ downloads and cover approximately $97.3\%$ of all \textit{Hugging Face} downloads.

Then, for each assessed task, we identified relevant benchmarks that assess model utility in relation to computational demands, particularly model size. The methodology details for analysing the AI benchmarks are described in 
\begin{natureReferences}
    Methods \ref{Methods:benchmarksMethodology}.
\end{natureReferences}
\begin{transactionsReferences}
    Appendix D,
    available in the online supplemental material.
\end{transactionsReferences}
 
A comprehensive list of the selected benchmarks and their corresponding tasks is presented in
\begin{natureReferences}
    Supplementary Information
\end{natureReferences} 
Table~\ref{tab:AIbenchmarks}, with further details provided in
\begin{natureReferences}
    Methods \ref{Methods:benchmarksDescription}.
\end{natureReferences}
\begin{transactionsReferences}
Appendix E,
    available in the online supplemental material.
\end{transactionsReferences}

\begin{allTabs}
    \begin{table*}[ht]
    \centering
    \renewcommand{\arraystretch}{1.3} 
    \begin{tabular}{|p{3.8cm}|p{3cm}|p{2.4cm}|p{1.8cm}|}
         \hline
         \textbf{Benchmark} & \textbf{AI Task (\textit{subtask, if applicable})}   & \textbf{Dataset} &  \textbf{Field} \\
         \hline
         
         OpenLLM Leaderboard \cite{open-llm-leaderboard-v1} & \multirow{4}{*}{Text Generation} & \multirow{5}{*}{HuggingFace}  & \multirow{6}{*}{Language}\\
         \cline{1-1}
         LMSys Chatbot Arena \cite{zheng2024judging} &  & & \\
         \cline{1-1}
         NPHardEval \cite{fan2023nphardeval} & & & \\
         \cline{1-1}
         BigCode Leaderboard \cite{bigcode-evaluation-harness} & & & \\
         \cline{1-2}

            mtebLeaderboard \cite{muennighoff2022mteb} & Text Classification &  &  \\
         \cline{1-3}
         
         WMT English-German \cite{bojar-etal-2014-findings} & Translation & Papers with Code  & \\
         \hline

        Open Object Detection Leaderboard \cite{open-od-leaderboard} & Object Detection & \multirow{2}{*}{Hugging Face} &  \multirow{3}{*}{Vision} \\
         
         \cline{1-2}
         Imagenet \cite{russakovsky2015imagenet} & Image Classification & & \\
         \cline{1-3}

         Semantic Segmentation on ADE20K \cite{Zhou_2017_CVPR} & Image Segmentation & Papers With Code & \\
         \hline
         
         Open ASR Leaderboard \cite{open-asr-leaderboard} & Automatic Speech Recognition & \multirow{2}{*}{Hugging Face} & \multirow{2}{*}{Audio} \\
         \cline{1-2}
         
         ARCH \cite{ARCH} & Audio Classification  & & \\
         \hline
         GenAI  \cite{ku2024imagenhub} & Text to Image  &  \multirow{2}{*}{HuggingFace} &  \multirow{2}{*}{Multimodal} \\

         \cline{1-2}
         MMMU Benchmark \cite{yue2023mmmu} & Image-Text to Text &  &  \\
         \hline
         
         Eth1-336 \cite{zhou2021informer} & Time Series Forecasting  & Papers with Code & Tabular\\
         \hline
    \end{tabular}
    \caption{\textbf{Evaluated AI Benchmarks}. AI Benchmarks evaluated in this study for analyzing the trade-off between model utility and size. For each benchmark, we also list also the corresponding task, source, and field.
    \label{tab:AIbenchmarks}
    }
\end{table*}

\end{allTabs}

\subsection{Utility dependency on model size}
\label{sec:AIbenchmarks:utilityOverModelSize}

\begin{figure*}
   \centering
   \includegraphics[width=0.5\textwidth]{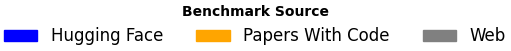}
   \hspace{1.0cm}
    \includegraphics[width=0.3\textwidth]{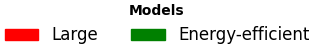}

    \normalsize{\textbf{Language}}
    \makebox[\textwidth][c]{
   \begin{subfigure}[]{0.5\textwidth}
   \begin{subfigure}[]{0.48\linewidth}
        \begin{subfigure}[]{\textwidth}
            \centering
            \includegraphics[width=\linewidth]{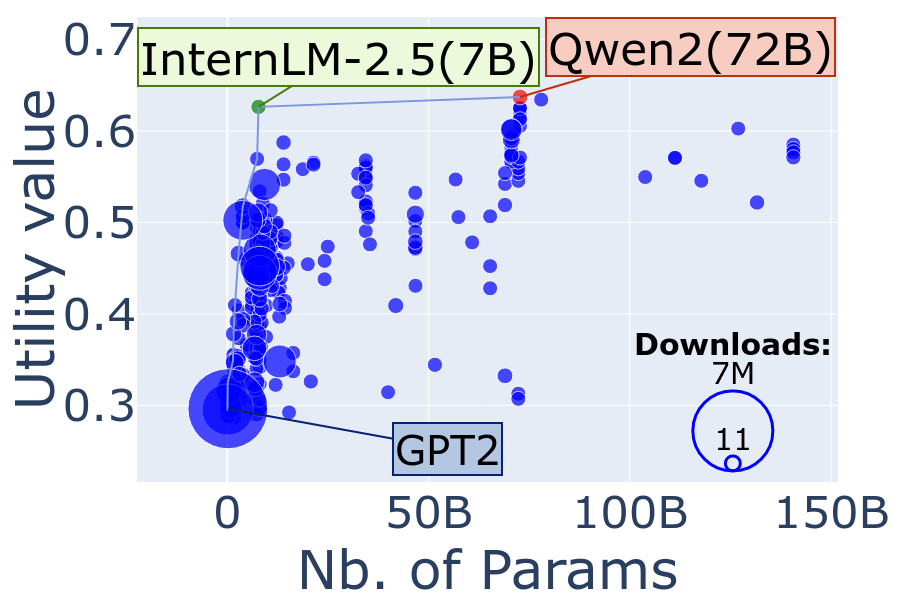}
            
            \scriptsize{\textit{(a1) OpenLLM}}
        \end{subfigure}
   \end{subfigure}
   \begin{subfigure}[]{0.48\linewidth}
        \begin{subfigure}[]{\textwidth}
            \centering
            \includegraphics[width=\linewidth]{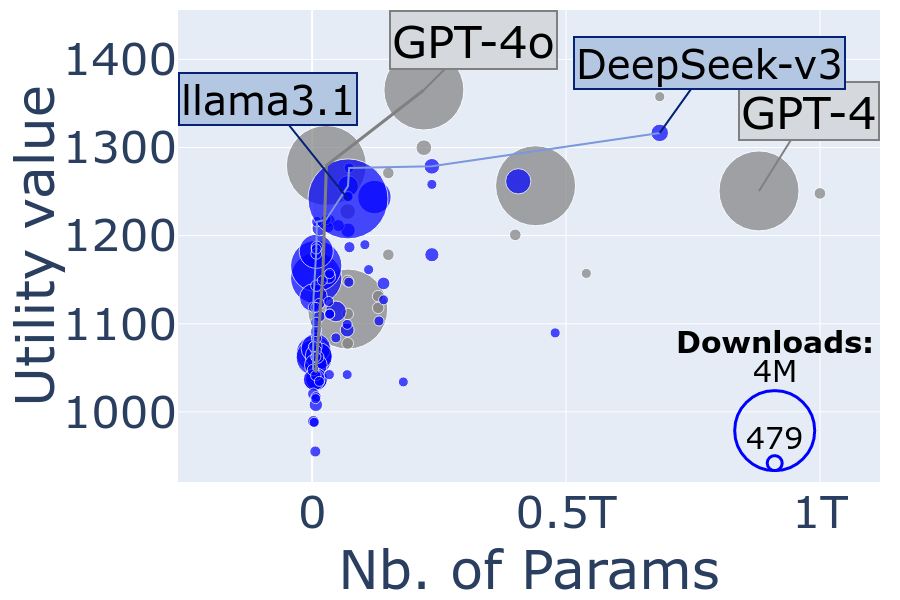} 

             \scriptsize{\textit{(a2) LMsys\label{fig:TotalAIbenchmarksAnalysis:LMsys}}}
        \end{subfigure}
    \end{subfigure}
    \caption{Text Generation}
    \label{fig:TotalAIbenchmarksAnalysis:text-generation}
    \end{subfigure}
   \begin{subfigure}[]{0.25\textwidth}
        \begin{subfigure}[]{\textwidth}
        \centering
         \includegraphics[width=\linewidth]{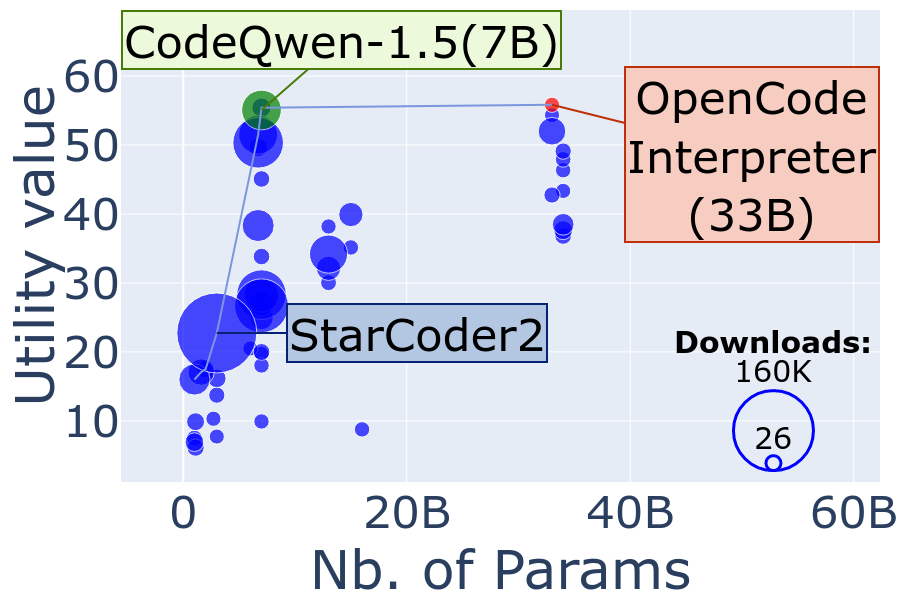} 
        \end{subfigure}
    \caption{Code Generation}
   \end{subfigure}
    \begin{subfigure}[]{0.25\textwidth}
        \includegraphics[width=\linewidth]{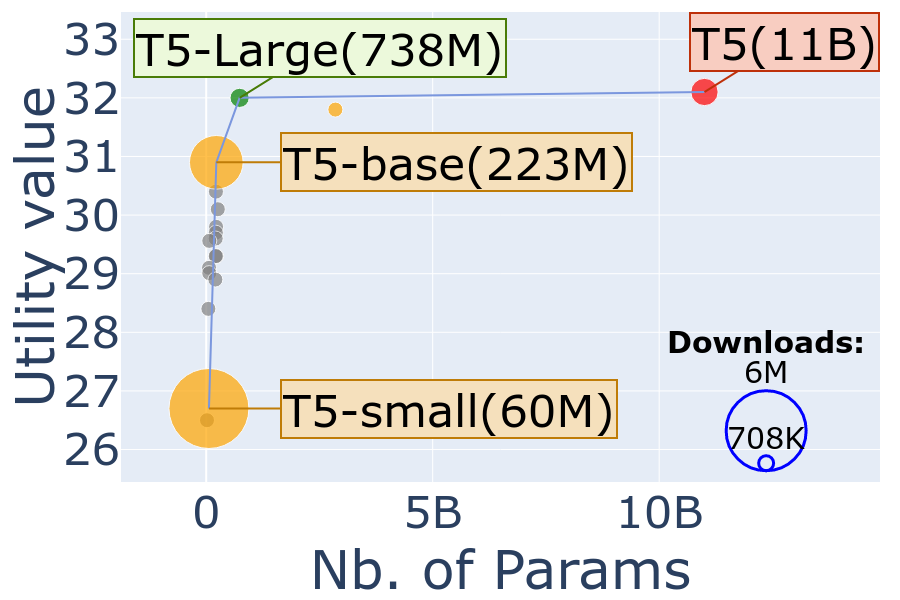}
        \caption{Translation}
   \end{subfigure}
   }
   \makebox[\textwidth][c]{
    \centering
   \begin{subfigure}[]{0.5\textwidth}
        \centering
        \begin{subfigure}[]{0.48\textwidth}
            \centering
             \includegraphics[width=\linewidth]{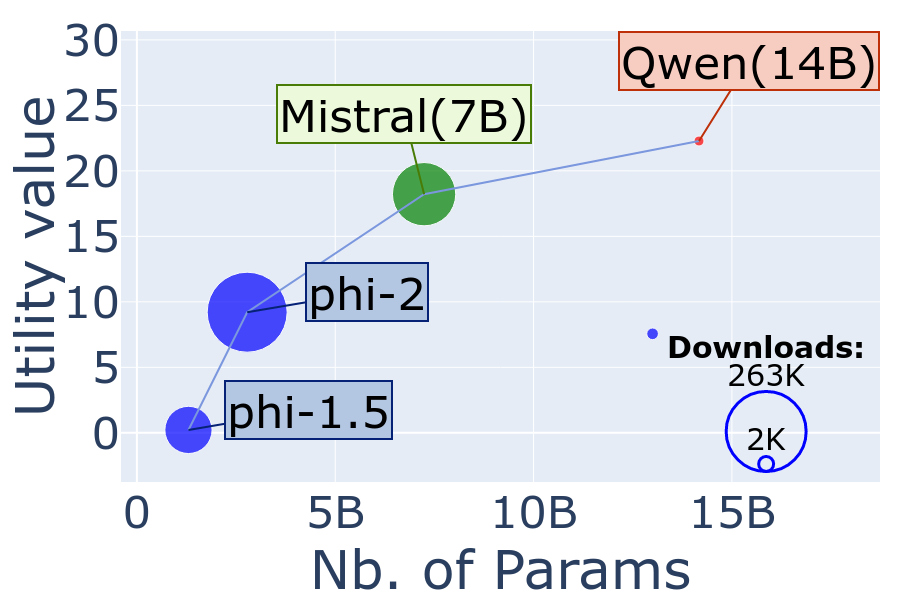}

             \scriptsize{(c1) \textit{Hugging Face} models}
        \end{subfigure}
        \begin{subfigure}[]{0.45\textwidth}
        \centering
        \includegraphics[width=\linewidth]{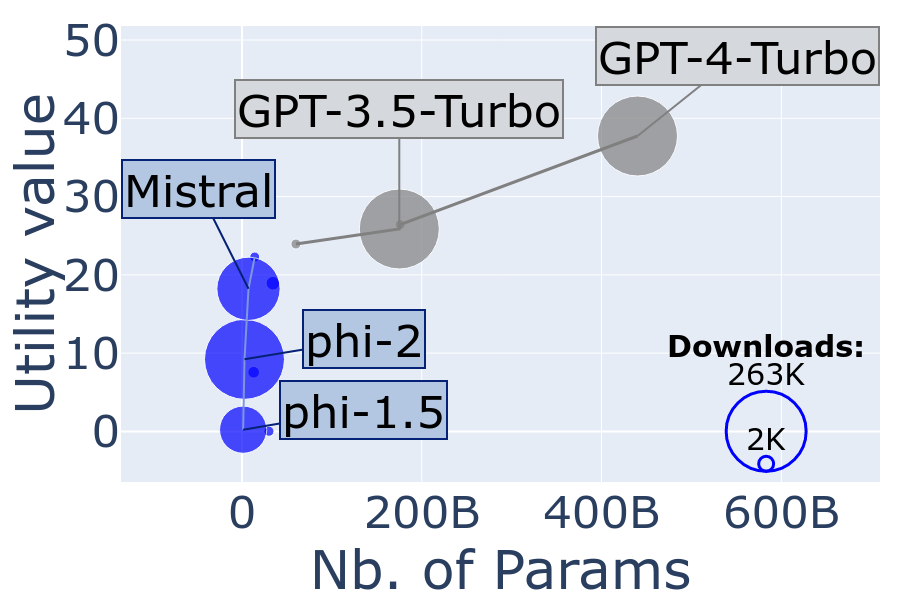}
        
        \scriptsize{(c2) \textit{Hugging Face} and large scale models}
        \end{subfigure}
        \caption{Mathematical Reasoning\label{fig:TotalAIbenchmarksAnalysis:Mathematical-Reasoning}}
   \end{subfigure}

   \begin{subfigure}[]{0.25\textwidth}
        \includegraphics[width=\linewidth]{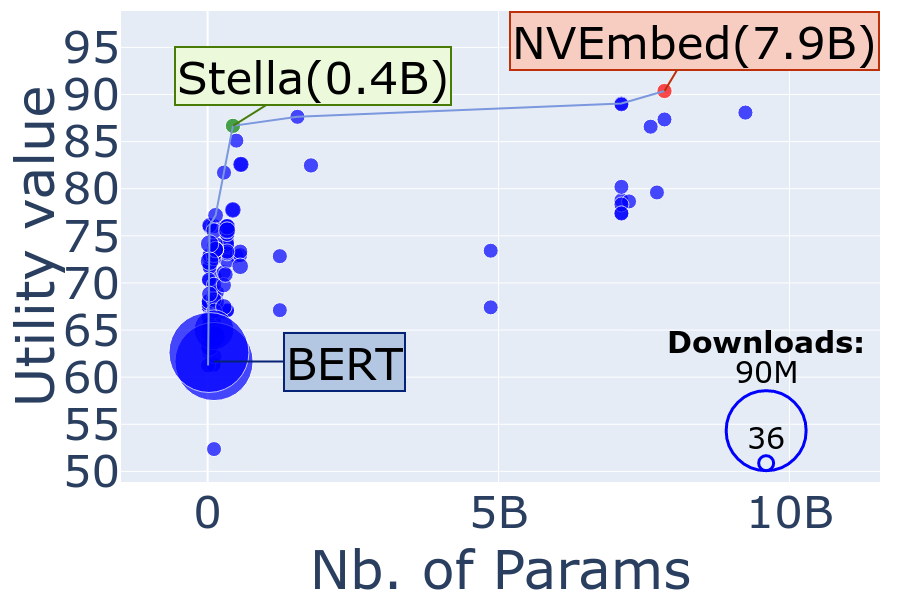} 
        \caption{Text Classification}
        \label{fig:TotalAIbenchmarksAnalysis:Text-Classification}
   \end{subfigure}

   \begin{subfigure}[]{0.25\textwidth}
        \includegraphics[width=\linewidth]{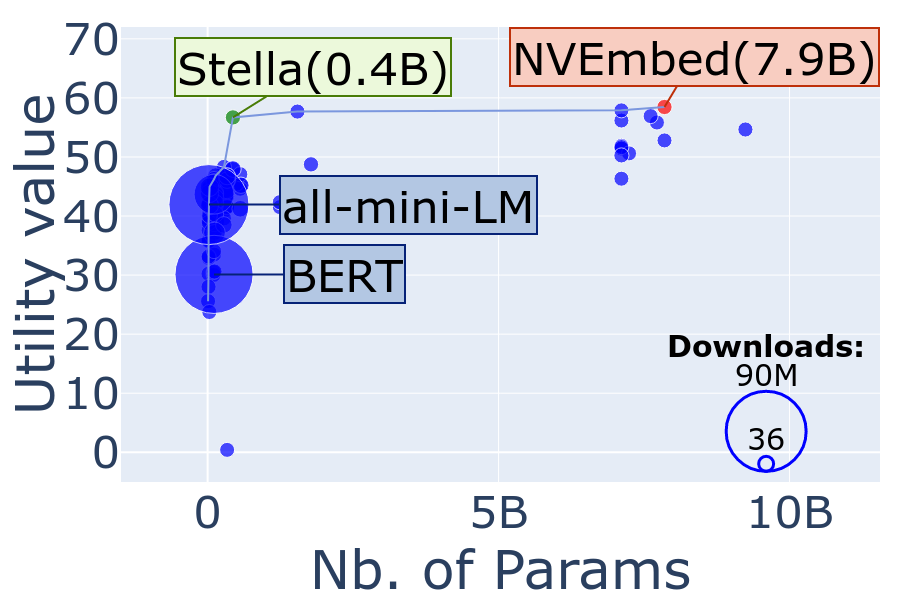} 
        \caption{Text Clustering}
        \label{fig:TotalAIbenchmarksAnalysis:Text-Clustering}
   \end{subfigure}
   
    }
    \makebox[\textwidth][c]{
    \begin{subfigure}[]{0.75\textwidth}
        \centering
        \normalsize{\textbf{Vision}}
        
        \begin{subfigure}[]{0.3\textwidth}
            \includegraphics[width=\linewidth]{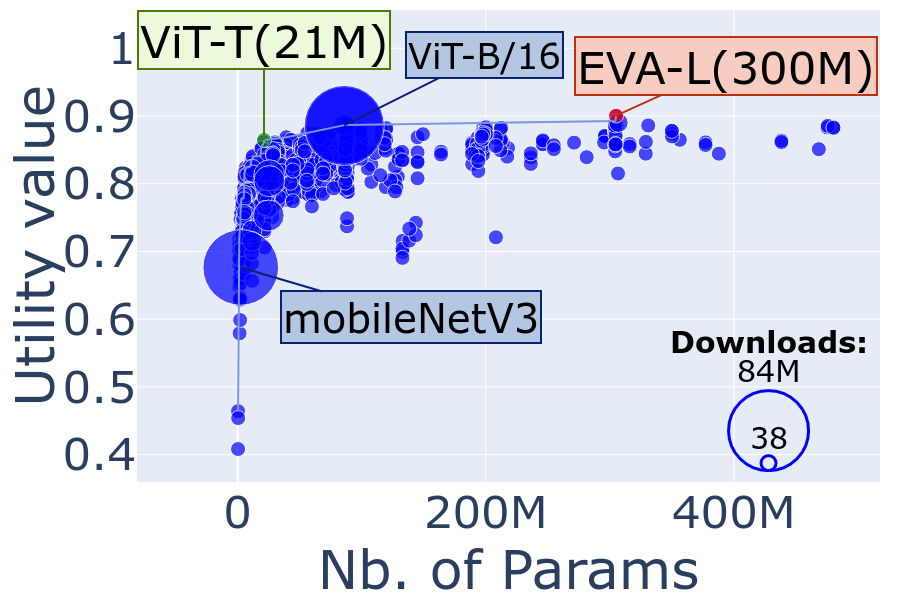}
            \caption{Image Classification}
            \label{fig:AIbenchmarks:image-classification}
       \end{subfigure}
       \begin{subfigure}[]{0.3\textwidth}
            \includegraphics[width=\linewidth]{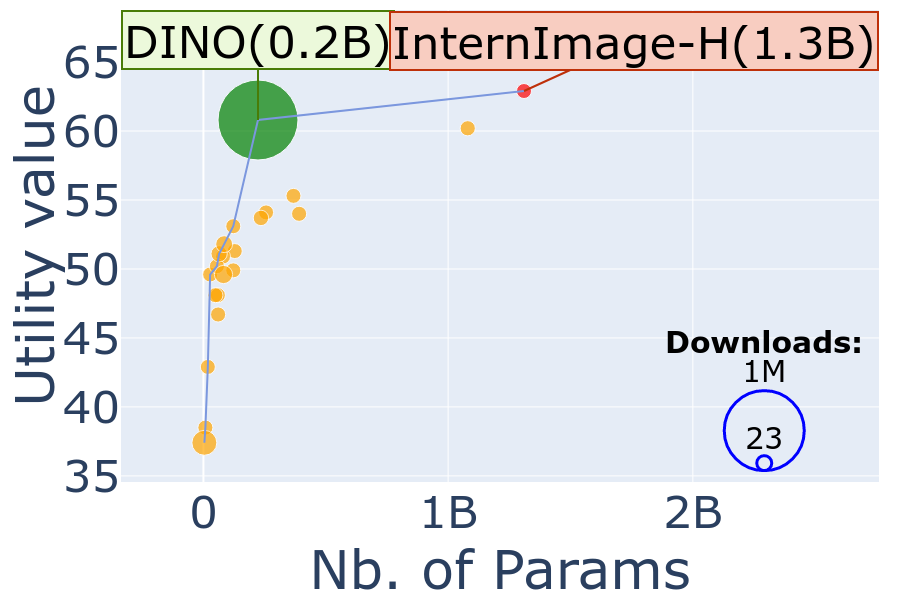} 
            \caption{Image Segmentation}
       \end{subfigure}
        \begin{subfigure}[]{0.3\textwidth}
            \includegraphics[width=\linewidth]{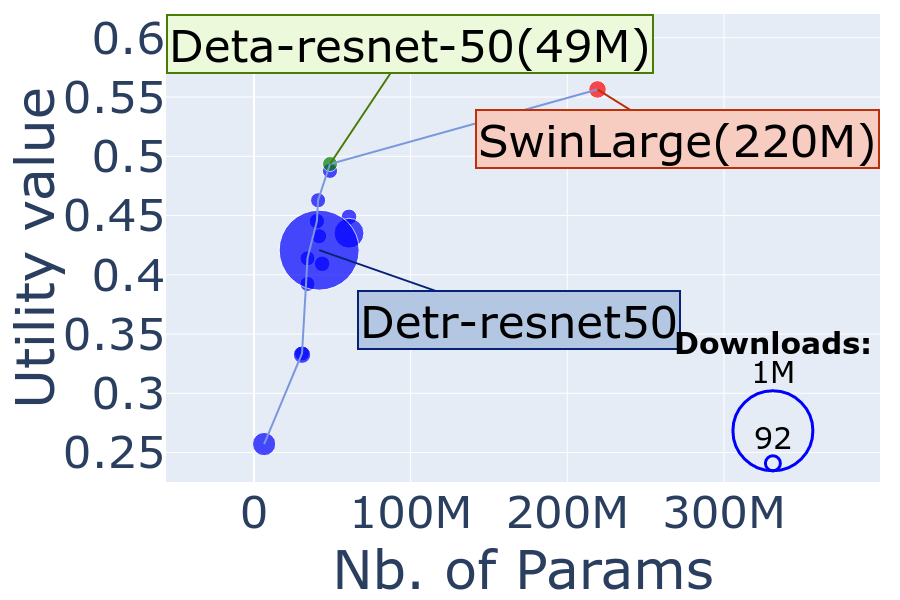} 
            \caption{Object Detection}
       \end{subfigure}
    \end{subfigure}

   \begin{subfigure}[]{0.25\textwidth}
        \centering
        \normalsize{\textbf{Tabular}}
        
        \includegraphics[width=\linewidth]{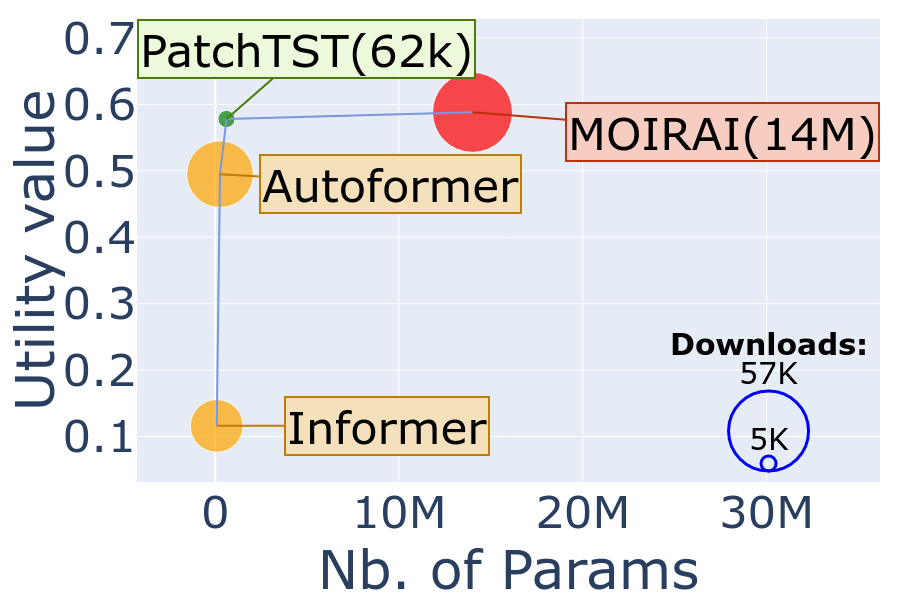} 
        \caption{Time Series Forecasting}
   \end{subfigure}
   }
   \makebox[\textwidth][c]{
   \begin{subfigure}[]{0.5\textwidth}
    \centering
    \normalsize{\textbf{Audio}}
    
        \begin{subfigure}[]{0.48\textwidth}
            \includegraphics[width=\linewidth]{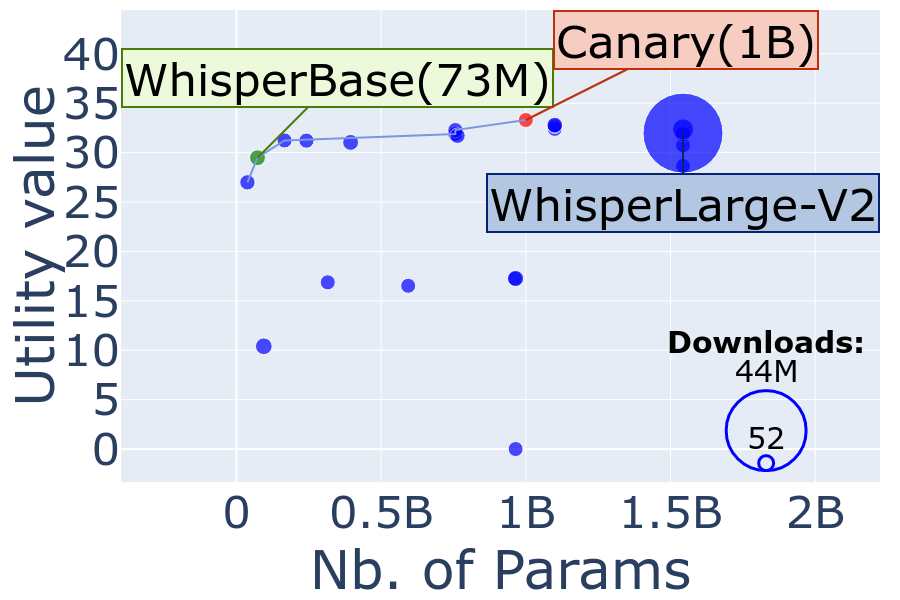}
            \caption{Speech Recognition}
            \label{fig:AIbenchmarks:speech-recognition}
        \end{subfigure}
        \begin{subfigure}[]{0.48\textwidth}
            \includegraphics[width=\linewidth]{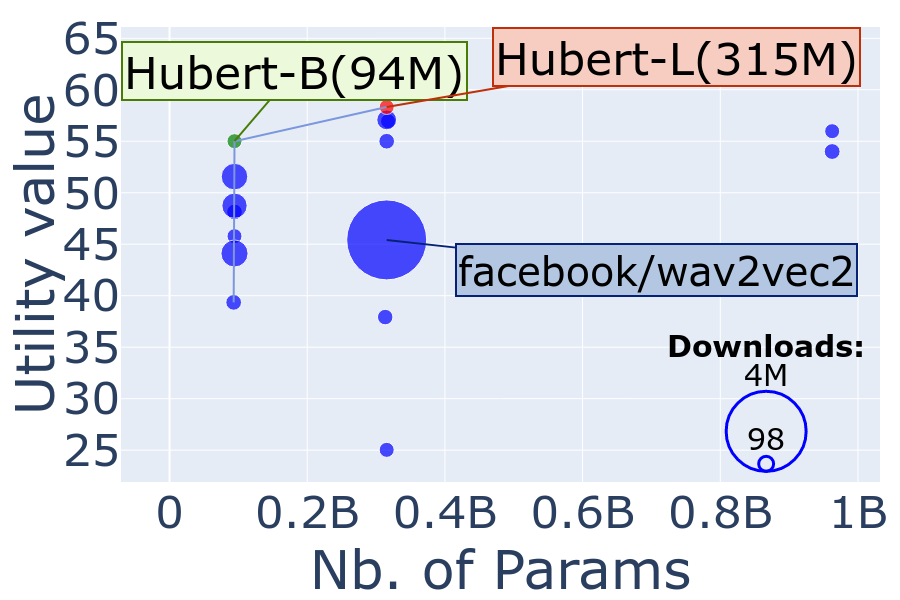} 
            \caption{Audio Classification}
        \end{subfigure}
    \end{subfigure}
    \begin{subfigure}[]{0.5\textwidth}
        \centering
        \normalsize{\textbf{Multimodal}}
        
         \begin{subfigure}[]{0.48\textwidth}
            \includegraphics[width=\linewidth]{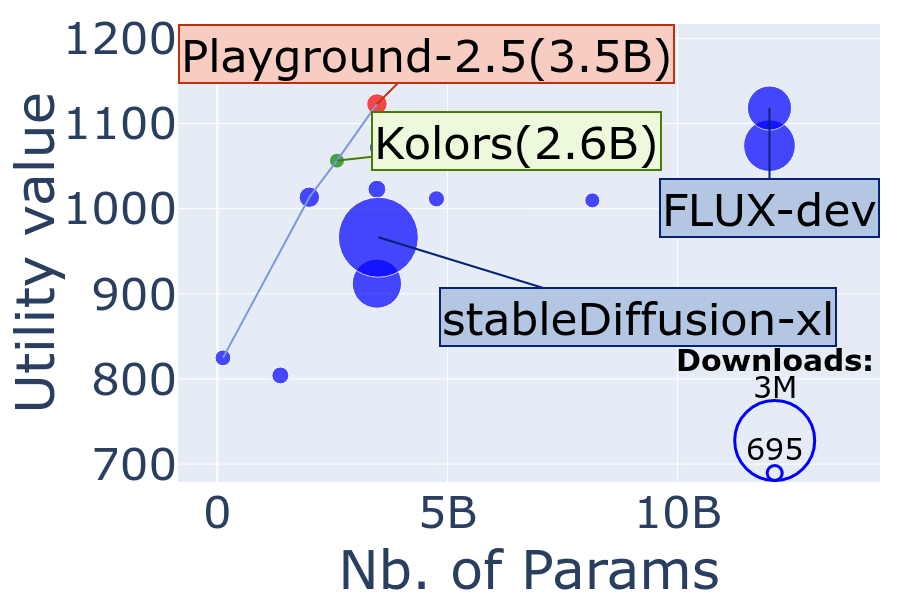}
            \caption{Text to Image\label{fig:TotalAIbenchmarksAnalysis:Text-to-Image}}
       \end{subfigure}
       \begin{subfigure}[]{0.48\textwidth}
            \includegraphics[width=\linewidth]{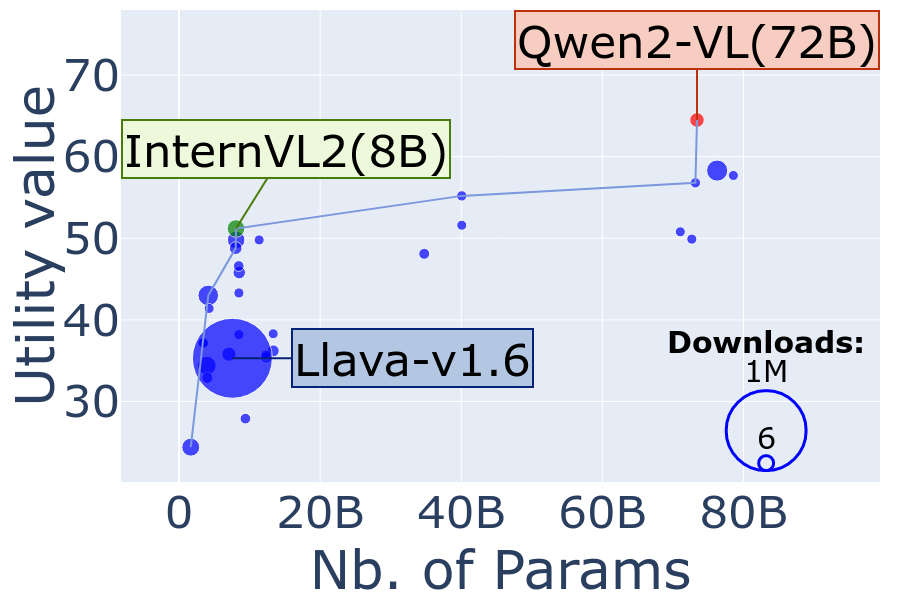} 
            \caption{Image-Text to Text \label{fig:TotalAIbenchmarksAnalysis:Image-Text-to-Text}}
       \end{subfigure}
    \end{subfigure}
   }
    \caption{\textbf{Tradeoff AI model utility vs size.} Model utility value versus number of parameters for popular AI tasks. Each point is a single model and its size is proportional to the model popularity. The green and red points indicate the \efficient and the \best models, respectively. An interactive version of the plots can be found in \url{https://tsb4.github.io/HF/}.}
    \label{fig:TotalAIbenchmarksAnalysis}
\end{figure*}

\begin{alternativeFigs}
    \begin{figure}
   \centering
   \centering
   \includegraphics[width=0.5\textwidth]{Figs/benchmarks/new_benchmarks/legend-benchmar-source.png}
   \hspace{1.0cm}
    \includegraphics[width=0.3\textwidth]{Figs/benchmarks/new_benchmarks/legend_models.png}

    \normalsize{\textbf{Language}}
    
   \begin{subfigure}[]{0.66\textwidth}
   \begin{subfigure}[]{0.48\textwidth}
        \begin{subfigure}[]{\textwidth}
            \centering
            \includegraphics[width=\linewidth]{Figs/benchmarks/new_benchmarks/acc_vs_params_openLLMLeaderboard.png}

            \scriptsize{\textit{(a1) OpenLLM}}
        \end{subfigure}
   \end{subfigure}
   \hfill
   \begin{subfigure}[]{0.48\textwidth}
        \begin{subfigure}[]{\textwidth}
            \centering
            \includegraphics[width=\linewidth]{Figs/benchmarks/new_benchmarks/acc_vs_params_lmsysLLM_bigLLMs_mixed.png} 

             \scriptsize{\textit{(a2) LMsys\label{fig:TotalAIbenchmarksAnalysis:LMsys}}}
        \end{subfigure}
    \end{subfigure}
    \caption{Text Generation}
    \label{fig:TotalAIbenchmarksAnalysis:text-generation}
    \end{subfigure}
    \hfill
   \begin{subfigure}[]{0.3\textwidth}
        \begin{subfigure}[]{\textwidth}
        \centering
        %
         \includegraphics[width=\linewidth]{Figs/benchmarks/new_benchmarks/acc_vs_params_big-code.png} 
        \end{subfigure}
    \caption{Code Generation}
   \end{subfigure}

    \centering
   \begin{subfigure}[]{0.8\textwidth}
        \centering
        \begin{subfigure}[]{0.45\textwidth}
            \centering
             \includegraphics[width=\linewidth]{Figs/benchmarks/new_benchmarks/acc_vs_params_NP_Hard_small.png}

             \scriptsize{(c1) \textit{Hugging Face} models}
        \end{subfigure}
        \hfill
        \begin{subfigure}[]{0.45\textwidth}
        \centering
        \includegraphics[width=\linewidth]{Figs/benchmarks/new_benchmarks/acc_vs_params_NP_Hard_mixed.png}
        \scriptsize{(c2) \textit{Hugging Face} and large scale models}
        \end{subfigure}
        \caption{Mathematical Reasoning\label{fig:TotalAIbenchmarksAnalysis:Mathematical-Reasoning}}
   \end{subfigure}
   
    
    \begin{subfigure}[]{0.3\textwidth}
        \includegraphics[width=\linewidth]{Figs/benchmarks/new_benchmarks/acc_vs_params_pwc_machine-translation-on-wmt2014-english-german.png}
        \caption{Translation}
   \end{subfigure}
   \hfill
   \begin{subfigure}[]{0.3\textwidth}
        \includegraphics[width=\linewidth]{Figs/benchmarks/new_benchmarks/acc_vs_params_mtebLeaderboard_class.png} 
        \caption{Text Classification}
   \end{subfigure}
   \hfill
   \begin{subfigure}[]{0.3\textwidth}
        \includegraphics[width=\linewidth]{Figs/benchmarks/new_benchmarks/acc_vs_params_mtebLeaderboard_clustering.png} 
        \caption{Text Clustering}
   \end{subfigure}

    \centering
    \normalsize{\textbf{Vision}}
    
    \begin{subfigure}[]{0.3\textwidth}
        \includegraphics[width=\linewidth]{Figs/benchmarks/new_benchmarks/acc_vs_params_merged-imagenet.png}
        \caption{Image Classification}
        \label{fig:AIbenchmarks:image-classification}
   \end{subfigure}
   \hfill
   \begin{subfigure}[]{0.3\textwidth}
        \includegraphics[width=\linewidth]{Figs/benchmarks/new_benchmarks/acc_vs_params_pwc_semantic-segmentation-on-ade20k.png} 
        \caption{Image Segmentation}
   \end{subfigure}
   \hfill
   \begin{subfigure}[]{0.3\textwidth}
        \includegraphics[width=\linewidth]{Figs/benchmarks/new_benchmarks/acc_vs_params_objectDetection.png} 
        \caption{Object Detection}
   \end{subfigure}
   
    \begin{subfigure}[]{0.6\textwidth}
    \centering
    \normalsize{\textbf{Audio}}
    
        \begin{subfigure}[]{0.48\textwidth}
            \includegraphics[width=\linewidth]{Figs/benchmarks/new_benchmarks/acc_vs_params_open-asr-leaderboard.png}
            \caption{Speech Recognition}
            \label{fig:AIbenchmarks:speech-recognition}
        \end{subfigure}
        \begin{subfigure}[]{0.48\textwidth}
            \includegraphics[width=\linewidth]{Figs/benchmarks/new_benchmarks/acc_vs_params_AudioClass.png} 
            \caption{Audio Classification}
        \end{subfigure}
    \end{subfigure}
    \hspace{1.0cm}
   \begin{subfigure}[]{0.3\textwidth}
        \centering
        \normalsize{\textbf{Tabular}}
        
        \includegraphics[width=\linewidth]{Figs/benchmarks/new_benchmarks/acc_vs_params_pwc_time-series-forecasting-on-etth1-336-1_fix.png} 
        \caption{Time Series Forecasting}
   \end{subfigure}

    \centering
    \normalsize{\textbf{Multimodal}}
    
     \begin{subfigure}[]{0.31\textwidth}
        \includegraphics[width=\linewidth]{Figs/benchmarks/new_benchmarks/acc_vs_params_genAI.png}
        \caption{Text to Image\label{fig:TotalAIbenchmarksAnalysis:Text-to-Image}}
   \end{subfigure}
   \hfill
   \begin{subfigure}[]{0.31\textwidth}
        \includegraphics[width=\linewidth]{Figs/benchmarks/new_benchmarks/acc_vs_params_MMMU_small.png} 
        \caption{Image-Text to Text \label{fig:TotalAIbenchmarksAnalysis:Image-Text-to-Text}}
   \end{subfigure}
   \hfill
   \begin{subfigure}[]{0.31\textwidth}
        \includegraphics[width=\linewidth]{Figs/benchmarks/new_benchmarks/acc_vs_params_MMMU_mixed.png} 
        \caption{Image-Text to Text (with large-scale LLMs)}
   \end{subfigure}
    \caption{Utility value over the number of parameters for several AI tasks. For each task evaluated, each point is a single model and the point size is proportional to its popularity. The points in green and red indicate the highlighted points, the \efficient and the \best, respectively.}
    \label{fig:TotalAIbenchmarksAnalysis}
\end{figure}
\joanna{in the second small points no unity is given for some values (no K or M, is it really this ? Only dozen of downloads ? ) and the difference of the size of the circles (between the big and small circle) seems strange in some figures ex: a1 - 7M and 11 (really 11, not 11k ?) and differences of sizes for c1 with 263K and 3K}
\end{alternativeFigs}

The selection of an AI model for inference within data centers is a critical factor influencing both energy consumption and the resulting utility. This section investigates how judicious model selection can promote energy efficiency without substantial compromise to utility. Specifically, we analyze the relationship between model size and utility across a range of benchmarks. Figure~\ref{fig:TotalAIbenchmarksAnalysis} presents this analysis, displaying the utility value of each evaluated model as a function of its parameter count. Each point on the graph corresponds to a distinct model within a given benchmark.
\footnote{All benchmarks are also available in \href{https://tsb4.github.io/HF/}{https://tsb4.github.io/HF/}} 
The size of each point is scaled to reflect model popularity, as measured by: (i) download counts for models hosted on \emph{Hugging Face} (blue); (ii) download counts of Hugging Face equivalent models for those hosted on \emph{Papers With Code} (orange); and (iii) web visit counts for models accessed through external APIs (gray)."


Within each benchmark, we highlight two key models. The {\em \efficient} model (green) balances high accuracy with a minimal parameter count, offering a favorable trade-off between performance and resource consumption. The {\em \efficient} model is selected as described in 
\begin{natureReferences}
    Methods~\ref{Methods:Keymodelsselection}.
\end{natureReferences}
\begin{transactionsReferences}
    Appendix D4
    available in the online supplemental material.
\end{transactionsReferences}
 Conversely, the {\best} model (red) is the one with maximum utility and typically has the highest number of parameters, often at the expense of increased energy usage.
The key models for each task are reported in
\begin{natureReferences}
    Supplementary Information Table \ref{tab:keyModels}.
\end{natureReferences}
\begin{transactionsReferences}
Table \ref{tab:keyModels}.
\end{transactionsReferences}


\begin{table*}[]
    \centering
    \begin{tabular}{|p{3cm}|p{5.3cm}|p{0.9cm}|p{0.9cm}|p{1.2cm}|p{1.2cm}|}
    \hline
    Task & Model & Params & Utility & Energy(J) & Downloads \\
    \hline
Text Generation & \textit{internlm/internlm2\_5-7b-chat (efficient)} & 8B & 0.6 & 8035.0 & 37281\\
\hline
Text Generation & \textbf{Qwen/Qwen2-72B-Instruct (best)} & 73B  & 0.6 & 35529.4* & 92091\\
\hline
Image Classification & \textit{timm/tiny\_vit\_21m\_512.dist\_in22k\_ft\_ in1k (efficient)} & 21M & 0.9 & 1907.0 & 1816\\
\hline
Image Classification & \textbf{timm/eva02\_large\_patch14\_448.mim \_m38m\_ft\_in22k\_in1k (best)} & 305M  & 0.9 & 5501.2 & 3201\\
\hline
Object Detection & \textit{jozhang97/deta-resnet-50-24-epochs (efficient)} & 49M & 0.5 & 4318.1 & 210\\
\hline
Object Detection & \textbf{jozhang97/deta-swin-large (best)} & 219M  & 0.6 & 8653.3 & 47849\\
\hline
Speech Recognition & \textit{openai/whisper-base.en (efficient)} & 73M & 29.5 & 724.3 & 605075\\
\hline
Speech Recognition & \textbf{nvidia/canary-1b (best)} & 1B  & 33.3 & 3726.0* & 11597\\
\hline
Image-Text to Text & \textit{OpenGVLab/InternVL2-8B (efficient)} & 8B & 51.2 & 84.4 & 126306\\
\hline
Image-Text to Text & \textbf{OpenGVLab/InternVL2-40B (best)} & 40B  & 55.2 & 298.7 & 5391\\
\hline
Text to Image & \textit{Kwai-Kolors/Kolors (efficient)} & 3B & 1056.4 & 2625.5 & 1914\\
\hline
Text to Image & \textbf{playgroundai/playground-v2.5-1024px-aesthetic (best)} & 3B  & 1123.0 & 3214.5 & 233322\\
\hline
Text Classification & \textit{NovaSearch/stella\_en\_400M\_v5 (efficient)} & 435M & 86.7 & 5824.7 & 385014\\
\hline
Text Classification & \textbf{nvidia/NV-Embed-v2 (best)} & 8B  & 90.4 & 12832.5 & 324552\\
\hline
Translation & \textit{google-t5/t5-large (efficient)} & 738M & 32.0 & 111.0 & 1028285\\
\hline
Translation & \textbf{google-t5/t5-11b (best)} & 11B  & 32.1 & 442.0 & 1648300\\
\hline
Audio Classification & \textit{ALM/hubert-base-audioset (efficient)} & 94M & 55.0 & 388.6 & 146\\
\hline
Audio Classification & \textbf{ALM/hubert-large-audioset (best)} & 315M  & 58.3 & 766.2 & 98\\
\hline
Image Segmentation & \textit{IDEA-Research/grounding-dino-base (efficient)} & 223M & 60.8 & 68.2 & 1064357\\
\hline
Image Segmentation & \textbf{OpenGVLab/internimage\_h\_22kto1k \_640 (best)} & 1B  & 62.9 & 175.2 & 23\\
\hline
Time Series Forecasting & \textit{ibm-granite/granite-timeseries-patchtst (efficient)} & 616K & 0.6 & 405.2 & 7030\\
\hline
Time Series Forecasting & \textbf{Salesforce/moirai-1.0-R-small (best)} & 14M  & 0.6 & 5606.5 & 56895\\
\hline
Code Generation & \textit{Qwen/CodeQwen1.5-7B-Chat (efficient)} & 7B & 55.1 & 7518.2 & 60861\\
\hline
Code Generation & \textbf{m-a-p/OpenCodeInterpreter-DS-33B (best)} & 33B  & 55.8 & 21035.5* & 507\\
\hline
Mathematical Reasoning & \textit{mistralai/Mistral-7B-Instruct-v0.1 (efficient)} & 7B & 18.2 & 7688.2* & 200741\\
\hline
Mathematical Reasoning & \textbf{Qwen/Qwen-14B-Chat (best)} & 14B  & 22.3 & 12004.3* & 2457\\
\hline
Text Clustering & \textit{NovaSearch/stella\_en\_400M\_v5 (efficient)} & 435M & 56.7 & 5824.7 & 385014\\
\hline
Text Clustering & \textbf{nvidia/NV-Embed-v2 (best)} & 8B  & 58.5 & 12832.5 & 324552\\
\hline
    \end{tabular}
    \caption{\textbf{Key models for each AI task.}
    \texttt{Energy-efficient} models are in \textit{italic} and \texttt{best-performing} models in \textbf{bold}.
    \label{tab:keyModels}
    }
\end{table*}

Across the benchmarks, the utility curves demonstrate a sublinear trend following the law of diminishing returns \cite{turgot}. Models with lower parameter counts exhibit a higher marginal gain in utility, reflecting their improved ability to capture data patterns. Conversely, larger models yield diminishing returns in utility with increasing parameter count 
(see
\begin{natureReferences}
    Supplementary Information
\end{natureReferences}
Figure~\ref{fig:dim-marginal-gain}).


\begin{figure*}[ht]
\centering

\begin{subfigure}[]{0.9\textwidth}
    \includegraphics[width=\textwidth]{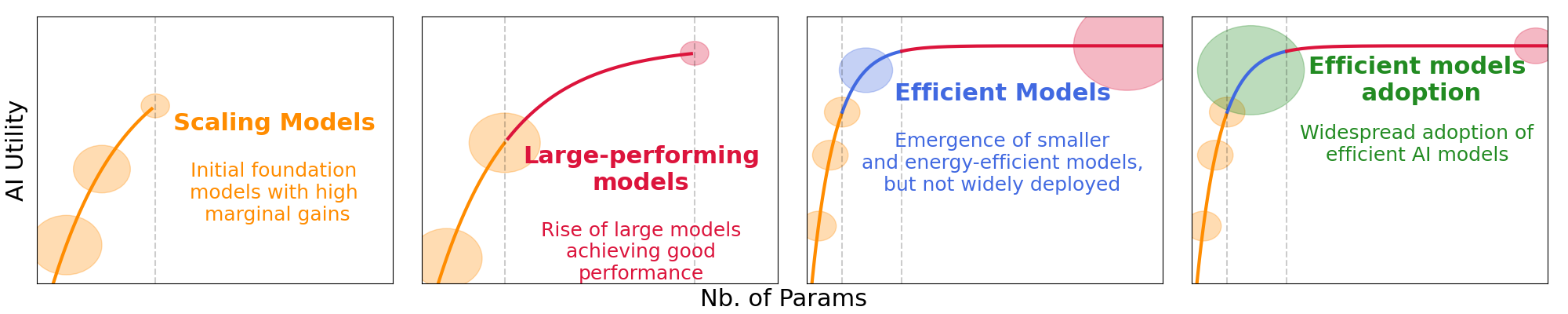}
    \caption{The 4 stages of development of an AI task over time \label{fig:4-stages}}
\end{subfigure}

\begin{subfigure}[]{0.9\textwidth}
    \includegraphics[width=\textwidth]{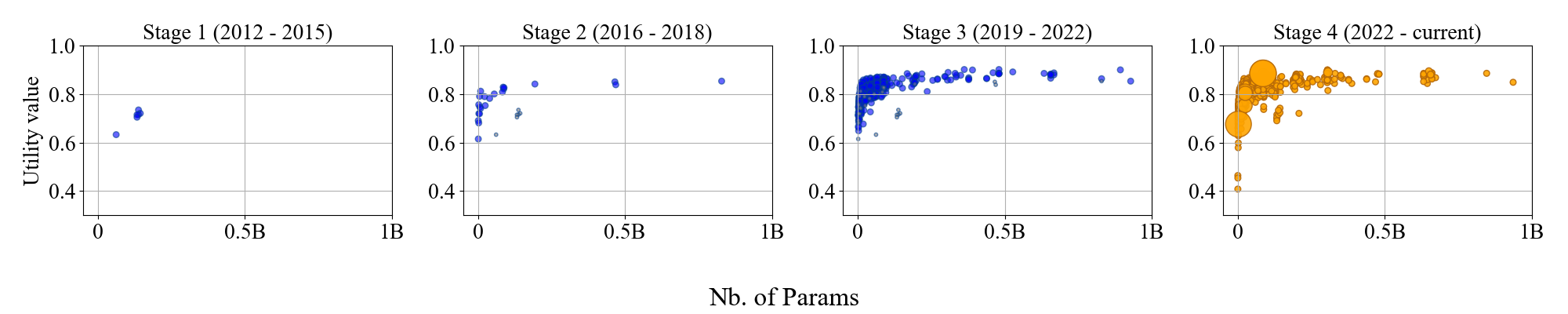}
    \caption{Example: \textit{Image Classification} \label{fig:4-stages-example}}
\end{subfigure}

\begin{subfigure}[]{0.9\textwidth}
    \includegraphics[width=\textwidth]{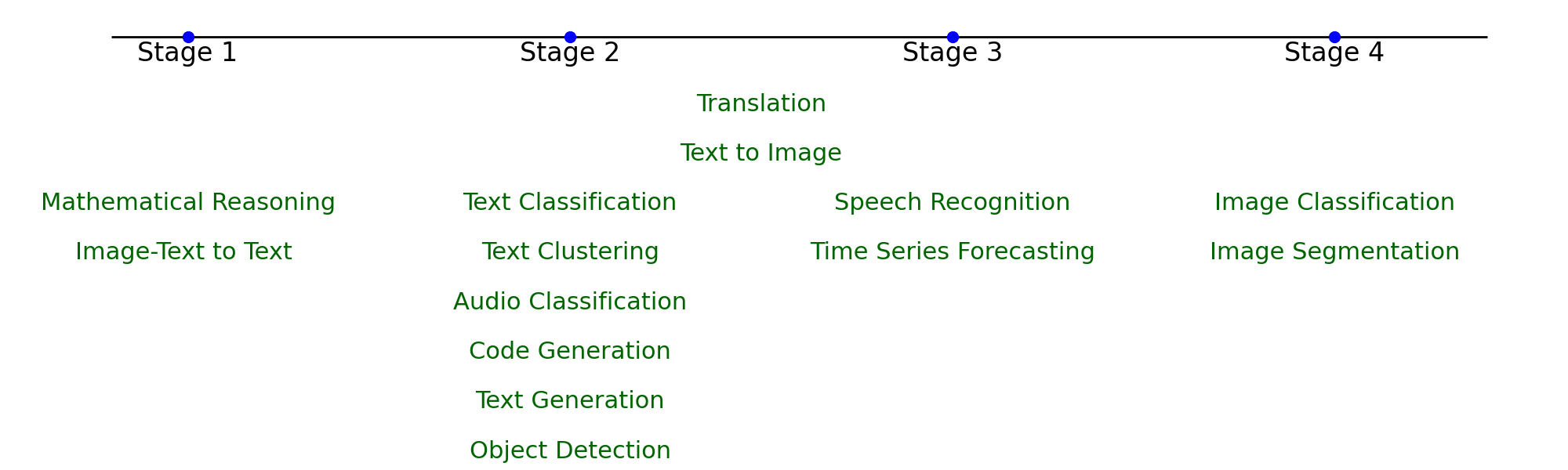}
    \caption{Stages of development of the most popular AI tasks  \label{fig:4-stages-all-tasks}}
\end{subfigure}
\caption{\textbf{The Four-Stage Development of AI Tasks: Balancing Accuracy and Model Size. }Top (\ref{fig:4-stages}). The four stages of development of an AI task, illustrating the evolving trade-off between accuracy and model size over time. Middle (\ref{fig:4-stages-example}) An example of these stages for \textit{Image Classification} Each point represent a model. The size of the orange circle in Stage 4 represents the model usage (number of downloads). Bottom (\ref{fig:4-stages-all-tasks}): Current development stages of major AI tasks. In Stage 1, the initial foundation models are developed.  In Stage 2, researchers focus on improving utility, leading to larger and performing models. In Stage 3, efficiency is introduced and smaller models with similar performance emerge. Last, in Stage 4, users partially adopt energy-efficient models. 
\label{fig:maturity-steps} }
\end{figure*}

\begin{figure}
\centering
\includegraphics[width=.8\linewidth]{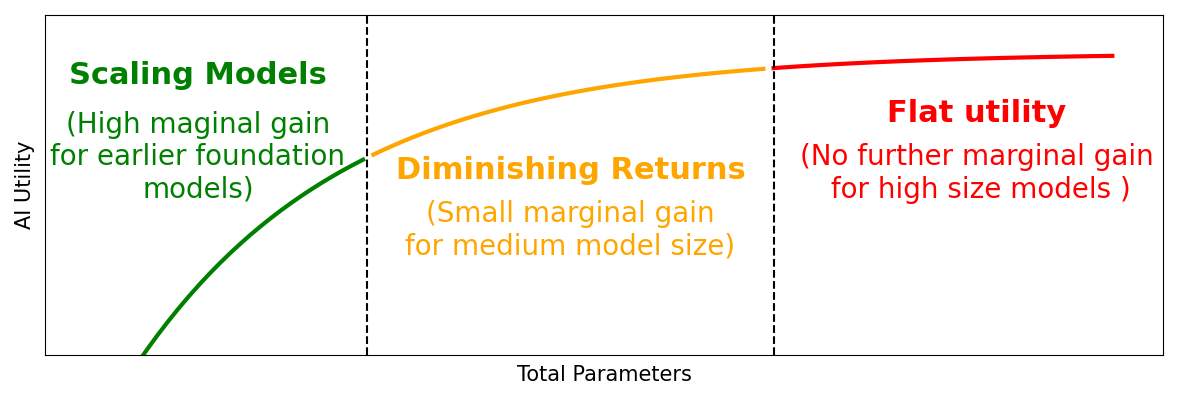}
\caption{\textbf{Law of diminishing return.} Diminishing Marginal Gains in AI Utility with Increasing Model Parameters. The relationship exhibits three phases: (i) Scaling Model: high initial marginal utility gain; (ii) Diminishing Returns: decreasing marginal gain with larger models; and (iii) Flat Utility: minimal marginal gain approaching a plateau. \label{fig:dim-marginal-gain}}
\end{figure}

This behavior aligns with scaling laws observed in language models \cite{kaplan2020scaling}, which posit a power-law relationship between model size, dataset size, and loss.  

This observation presents a significant opportunity for enhancing energy efficiency in AI inference. By transitioning from large, high-performing models to smaller, energy-efficient alternatives, substantial reductions in model size and, consequently, energy consumption can be achieved with a minimal impact on utility. For example, in \textit{speech recognition}, the best-performing model is 14 times larger than the energy-efficient model, while providing only a 7.8\% improvement in utility.

The maturity of an AI task significantly influences the utility-size trade-off. As a task matures with a high-performing model, subsequent development often focuses on creating smaller, yet comparably effective, models \cite{desislavov2023trends}. 
This trend is particularly evident in mature and well-established domains such as \textit{image classification}, \textit{audio classification}, 
and \textit{speech recognition}. 
For these tasks, extensive research has optimized model efficiency over time. Consequently, the Pareto frontier relating model size to utility is almost flat for large models, indicating a very small increase in utility with increasing number of parameters.

\subsection{Model Adoption}
\label{sec:model-adoption}

Model usage patterns play an important role in responsible AI. As exemplified in Figure~\ref{fig:TotalAIbenchmarksAnalysis}, all users do not use the same model to solve the same task. Even more, a large number of them do not use the best available models. In fact, model adoption is influenced by several factors: model and brand popularity, habit~\cite{labrecque2017habit,heidenreich2016innovations,ram1989consumer}, model size relative to available hardware, developer vs. general public usage, task maturity, and so on. 

First, due to brand popularity and habit, the most widely used models (big blue dot) tend to be models from famous model families: {\tt GPT2}, {\tt GPT4} for LLM, {\tt Phi2}, {\tt Mistral}, {\tt BERT} for \textit{Mathematical Reasoning}, {\tt StableDiffusion} for \textit{Text to Image}, even if better alternatives exist for some tasks (see e.g.,~Figure~\ref{fig:TotalAIbenchmarksAnalysis:Text-to-Image}). 

Second, due to the price and availability of hardware, especially the best GPUs, model size has a large impact on model usage by developers. For \textit{Text Generation}, the \textit{OpenLLM benchmark}, which considers models available on the \textit{Hugging Face} platform, suggests that some small models (e.g., \texttt{GPT-2}) tend to be very popular (with large numbers of downloads) among developers due to hardware limitations that limit the adoption of large models. Similarly, {\tt BERT} is very popular for \textit{Text Classification} and \textit{Text Clustering}. 

However, the general public has access to very large models through web applications and APIs. This is especially true for the \textit{Text Generation} task, where users have access to large language models with hundreds of billions of parameters. The \textit{LMsys benchmark} in Figure~\ref{fig:TotalAIbenchmarksAnalysis:text-generation} shows the very high usage of such models. Notably, models within the \texttt{GPT-4} family attract over three billion visits per month,
according to SimilarWeb \cite{similarweb2025},
demonstrating the significant impact of these powerful systems.

\begin{alternativeFigs}
    \begin{figure}
\centering
\includegraphics[width=.8\textwidth]{Figs/board.pdf}
\caption{Maturity Steps \label{maturity-steps}}
\end{figure}

\begin{figure}
\centering
\includegraphics[width=.6\textwidth]{Figs/story/Maturity_sketches.png}
\caption{Evolution of AI efficiency, showing the relationship between utility and model size (number of parameters). Blue points mark the main usage patterns at each stage: In Stage 1, the performing models are initially created; In Stage 2, researchers focuses on improving utility, leading to larger models; In Stage 3,efficiency is introduced and smaller models with similar performance; and in Stage 4, the "small is sufficient" paradigm is achieved, where efficient and high-performing models are widely utilized. \ramon{Increase the ball size corresponding to the popular model.} \tiago{To be removed} \label{maturity-steps}}
\end{figure}

\end{alternativeFigs}

The task maturity also influences the size of adopted models. Its evolution over time can be divided into 4 main phases (see Figure~\ref{fig:maturity-steps}).
In the first phase, the task is not mature and high marginal gains are experienced with new larger models. During this phase, a large number of users use models with performance very far from state-of-the-art models, as the field is swiftly evolving. A typical example is reasoning with the two benchmarks \textit{Mathematical Reasoning} (see Figure~\ref{fig:TotalAIbenchmarksAnalysis:Mathematical-Reasoning}) and \textit{Image-Text to Text} (college-level questions), see Figure~\ref{fig:TotalAIbenchmarksAnalysis:Image-Text-to-Text}.

In the second phase, we experience the apparition of large performing models which are not yet massively used by the community. \textit{Text Classification} (see Figure~\ref{fig:TotalAIbenchmarksAnalysis:Text-Classification}) and \textit{Text Clustering} (see Figure~\ref{fig:TotalAIbenchmarksAnalysis:Text-Clustering}) have very performing and large models, e.g., the state of the art {\tt NVEmbed} model. However, users are massively using a {\tt BERT} model, with a utility 30\% lower. 

In the third phase, the task is mature. The marginal gain is now small. The \best model is adopted (if not too large) and the community introduces small efficient models in parallel. \textit{Speech recognition} (= \textit{Audio-to-text}) is typical of such a phase (see Figure~\ref{fig:AIbenchmarks:speech-recognition}). A large performing model, {\tt WhisperLarge-V2}, is mostly used. Small efficient models have been developed, as the \efficient model, {\tt WhisperBase}.

Other tasks are in transition between phases 2 and 3. 
One example is \textit{Text to Image} (see Figure~\ref{fig:TotalAIbenchmarksAnalysis:Text-to-Image}): while high-performing models like {\tt Flux-dev} have been proposed and gained significant adoption, lower-performing models such as {\tt stable-diffusion-xl} remain widely used, likely due to adoption barriers. Efficient smaller models like {\tt Playground-2.5} and {\tt Kolors} have emerged but have not yet achieved widespread use.




The last phase corresponds to very mature tasks for 
which performing and small models have been developed and adopted. The share of usage between \efficient and \best models depends on model sizes. If the latter are not too big, both models will be used. 
The typical example is \textit{Image Classification} (see Figure~\ref{fig:AIbenchmarks:image-classification}). For this task, efficient models , such as \texttt{MobileNetV3} and \texttt{ViT-T}, developed by \textit{Google}, are widely adopted by users.



The maturity level of each AI task, the size of its models, their adoption pattern, the existence or not of small and efficient models will thus have a direct impact on how much energy savings can be achieved through model selection. 
In the next section, we explore the potential energy savings of each of these tasks using model selection.

\section{Estimating the Savings of AI Model Selection}
\label{sec:energyEstimation}

This section quantifies the energy savings achievable through model selection. We first estimate the energy consumption of the benchmarked AI models and then analyze the energy reductions resulting from applying model selection techniques.


\subsection{AI Inference Energy Consumption Measurement Methodology}
\label{sec:measurementmethodology}

\begin{allFigs}
    \begin{figure*}[ht]
    \centering
    \begin{subfigure}[]{\textwidth}
    \includegraphics[width=\linewidth]{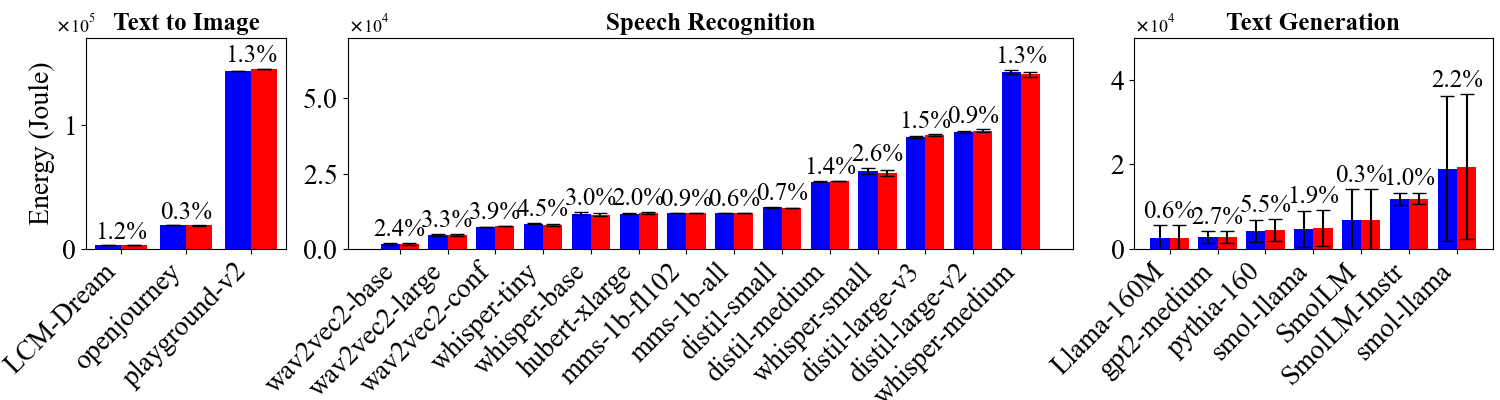}
    \vspace{0.5cm}
    \includegraphics[width=\linewidth]{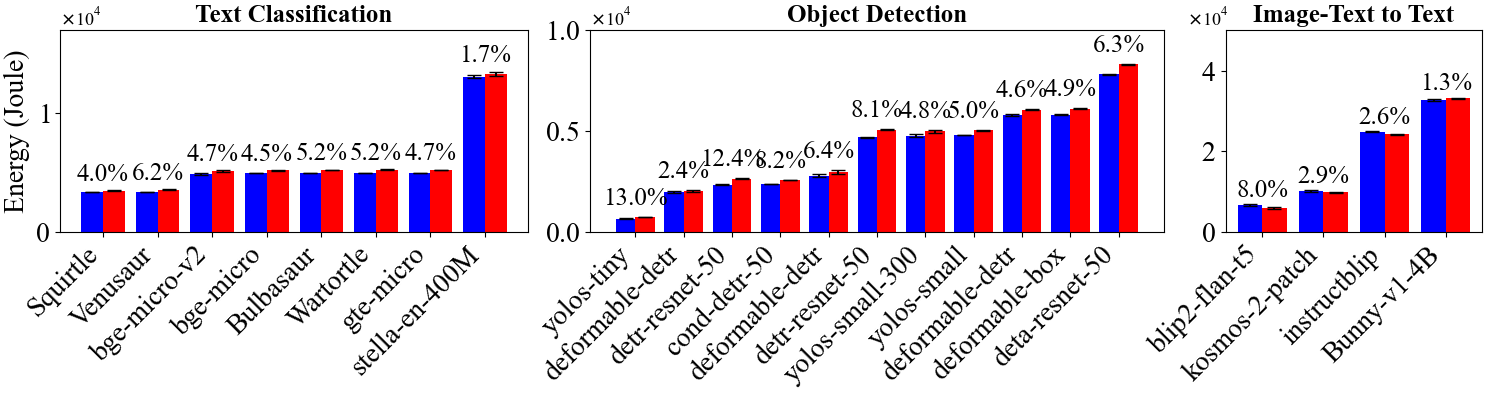}
    \end{subfigure}
    \caption{\textbf{Validation of the \emph{CabonTracker} tool.} Comparison of energy consumption measurements obtained using the software-based tool \emph{CabonTracker} (CT) and the \emph{PowerSpy} power meter device, across models from several tasks. In blue, we show the \textit{CabonTracker} measurements and in red, the \textit{PowerSpy} measurements.
    \label{fig:energyConsumption:comparisonCTandPowerspy}
    }
\end{figure*}
\end{allFigs}

Precise measurements of the energy consumption are crucial for selecting energy-efficient models during inference.
Several works \cite{luccioni2023estimating, luccioni2024power, yang2024harnessing, dodge2022measuring} have proposed the energy monitoring of AI models in order to identify opportunities for enhancing energy efficiency. These works usually use specialized software-based tools for measuring the models energy consumption, such as CarbonTracker \cite{anthony2020carbontracker}, CodeCarbon \cite{benoit_courty_2024_11171501}, Zeus \cite{zeus:nsdi23}, and Scaphandre \cite{scaphandre}. We opted for \textit{CarbonTracker}, which monitors CPU and GPU energy consumption using built-in sensors, since it is well documented and widely used by researchers. Further details are described in 
\begin{natureReferences}
    Methods~\ref{sec:methods:carbonTracker}.
\end{natureReferences}
\begin{transactionsReferences}
    Appendix A,
    available in the online supplemental material.
\end{transactionsReferences}

In this section, we present a methodology for assessing the energy consumption of AI models based on the usage of \textit{CarbonTracker}. 

We validated the tool by comparing its measurements with those of \textit{PowerSpy} \cite{powerspy2}, a hardware-based power meter that can assess the power consumption of a computer. The experimental setup is described in 
\begin{natureReferences}
    Methods \ref{sec:methods:experimentalSetupCarbonTracker}.
\end{natureReferences}
\begin{transactionsReferences}
    Appendix B,
    available in the online supplemental material.
\end{transactionsReferences}

\begin{natureReferences}
    Supplementary Information
\end{natureReferences}

The results of our validation are described in Figure~\ref{fig:energyConsumption:comparisonCTandPowerspy}. They revealed a small average difference of 3.42\% for the software-based \textit{CarbonTracker} compared to \emph{PowerSpy}. This close agreement establishes \textit{CarbonTracker} as a reliable tool for energy measurement in our experiments.

Given the large number of models per AI task, 
it was impractical to measure the energy consumption of each one directly. In addition, some models were too large to run. Therefore, for each task, we measured the energy consumption of \efficient and \best models when feasible. For the remaining models, we estimated their energy consumption using a power-law function, as suggested in the literature~\cite{desislavov2023trends}. 

We investigated the relationship between model size (in number of parameters) and energy consumption across three AI tasks: image classification, speech recognition, and text generation. The energy consumption was measured using CarbonTracker (see 
\begin{natureReferences}
    Methods \ref{sec:methods:experimentalSetupNumberOfParams}).
\end{natureReferences}
\begin{transactionsReferences}
    Appendix C,
    available in the online supplemental material).
\end{transactionsReferences}

Figure~\ref{fig:energyMethodology:energy_vs_params} shows the median energy consumption over 10 inference requests per model as a function of model size. Across all tasks, energy consumption scales linearly with model size on a log-log scale. Then, we can approximate by a polynomial relationship on the linear scale (see 
\begin{natureReferences}
    Methods \ref{sec:methods:energyEstimationExperimentalSetup}).
\end{natureReferences}
\begin{transactionsReferences}
    Appendix F,
    available in the online supplemental material).
\end{transactionsReferences}

\begin{figure}
    \centering
    
    



     \includegraphics[width=\linewidth]{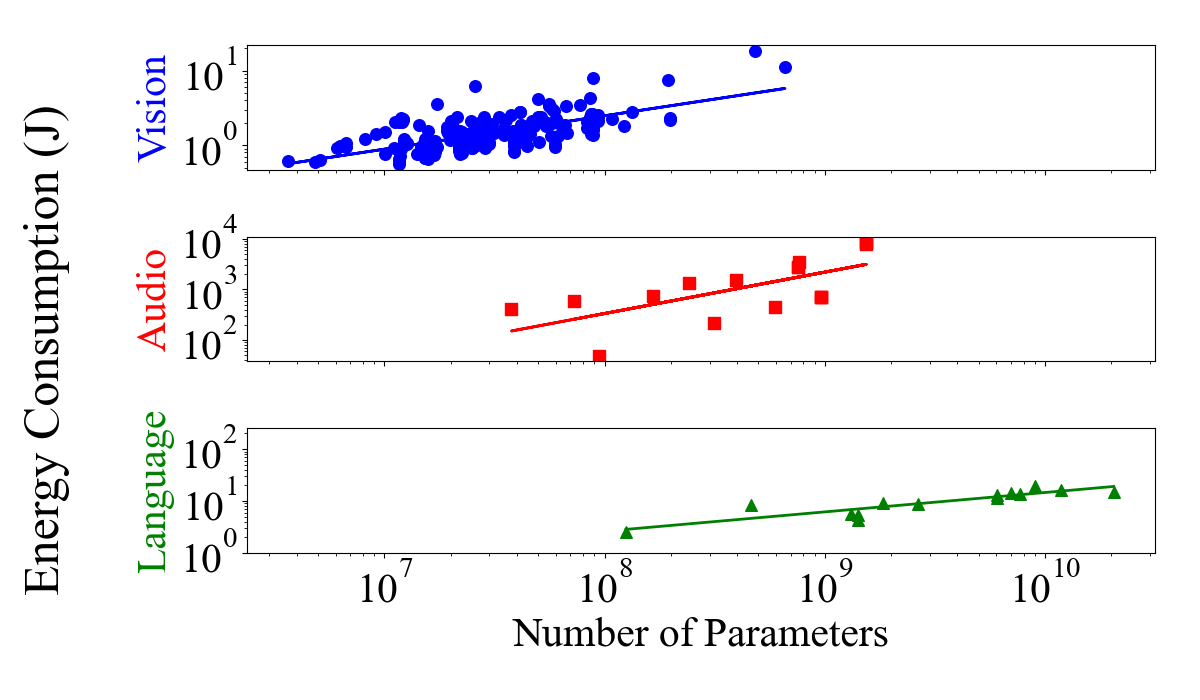}%
    \label{fig:energyMethodology:LLM:energy_vs_params}
    
    \caption{\textbf{Modeling AI model energy consumption.} Inference energy consumption versus model size (number of parameters) for \textit{Image Classification} (vision), \textit{Speech Recognition} (audio), and \textit{Text Generation} (language) tasks. Median energy values from 10 repeated inference requests are plotted for each data point.}
    \label{fig:energyMethodology:energy_vs_params}
\end{figure}

This result is consistent with Desislavov et al.~\cite{desislavov2023trends}, who reported a linear relationship between the number of parameters and floating-point operations on a log-log scale for vision models, and a linear scaling of energy consumption with the number of operations.

Thus, model size can be usd for estimating inference energy consumption when direct measurement is impractical, e.g., when the model is too large.

\subsection{Savings by Model Selection}
\label{sec:savings}

Equipped with our energy measurements and estimates, we can now evaluate how much energy can be saved through model selection for all the AI tasks under consideration. 

We first discuss our choice of the \efficient (green) model for each task by comparing it to the \best available (red) model for that task. 
The methodology used for selecting the key models is described in
\begin{natureReferences}
    Methods \ref{Methods:Keymodelsselection} and the models are listed
\end{natureReferences}
\begin{transactionsReferences}
    Appendix D4,
    available in the online supplemental material
    and the models are listed in Table~\ref{tab:keyModels}.
\end{transactionsReferences}

\begin{natureReferences}
    in Supplementary Information Table \ref{tab:keyModels}.
\end{natureReferences}
\begin{transactionsReferences}
\end{transactionsReferences}

Figure~\ref{fig:energySobrietyPotential:scenario1} shows the impact of switching from the \best model to the \efficient model for each task. On average, the latter is 65.8\% more energy efficient than the former, at the small cost of losing only 3.9\% of utility. 
If we distinguish between tasks, the savings are around 70\% or more for mature tasks such as \textit{Time Series Forecasting} (92.8\%), \textit{Speech Recognition} (80.6\%) or \textit{Image Classification} (69.6\%), while they are lower for immature tasks such as \textit{Mathematical Reasoning} (36\%) and \textit{Text-to-Image} 
(image generation) (only 18.3\%). The large savings with limited impact on the utility validate the choice of models.

\begin{allFigs}
   \begin{figure}
    \centering
    \includegraphics[width=\linewidth]{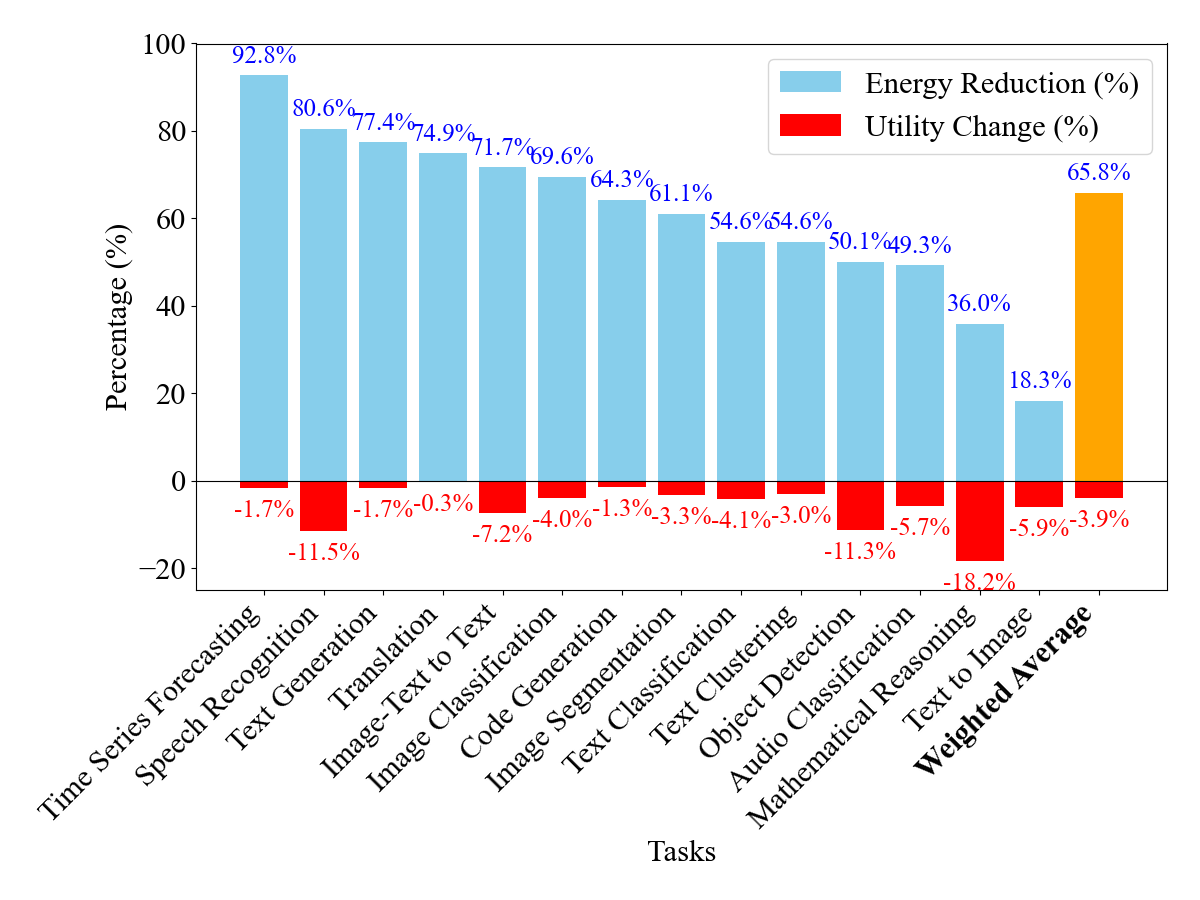}
    \caption{\textbf{Energy and Utility Impact of Switching to Efficient Inference Models. }This figure presents the percentage difference in energy reduction and utility when shifting AI inference from the \best model to the \efficient model for various tasks. The rightmost bar (i.e., the bar in orange and red) represents the weighted average across all considered AI tasks, with weights based on their usage (number of downloads). }
    \label{fig:energySobrietyPotential:scenario1}
\end{figure}

\end{allFigs}

Users may have different tolerances for utility loss depending on their specific needs. To account for this, we evaluated global energy savings and utility variations under different maximum utility drop values when selecting the \efficient model. As expected, allowing for larger utility drop results in greater energy savings with the cost of reduced global utility. Our findings (see 
\begin{natureReferences}
    Supplementary Information 
\end{natureReferences}
Figure \ref{fig:energySavingsAccDropScenario1}) indicate that global energy savings can reach up to 88.3\% when a maximum global utility loss of 37.5\%.

We now take into account the model's usage. We consider a scenario in which all users using a larger model than the \efficient (green) model of the task switch to the latter model. We examine the energy savings, as well as the impact on the  utility. 
Figure~\ref{fig:energySobrietyPotential:scenario3} illustrates the estimated percentage reduction in energy consumption for each task and the average reduction for all the tasks weighted
by the total number of downloads per task on \textit{Hugging Face}.

\begin{figure}[ht]
    \centering
        \includegraphics[width=\linewidth]{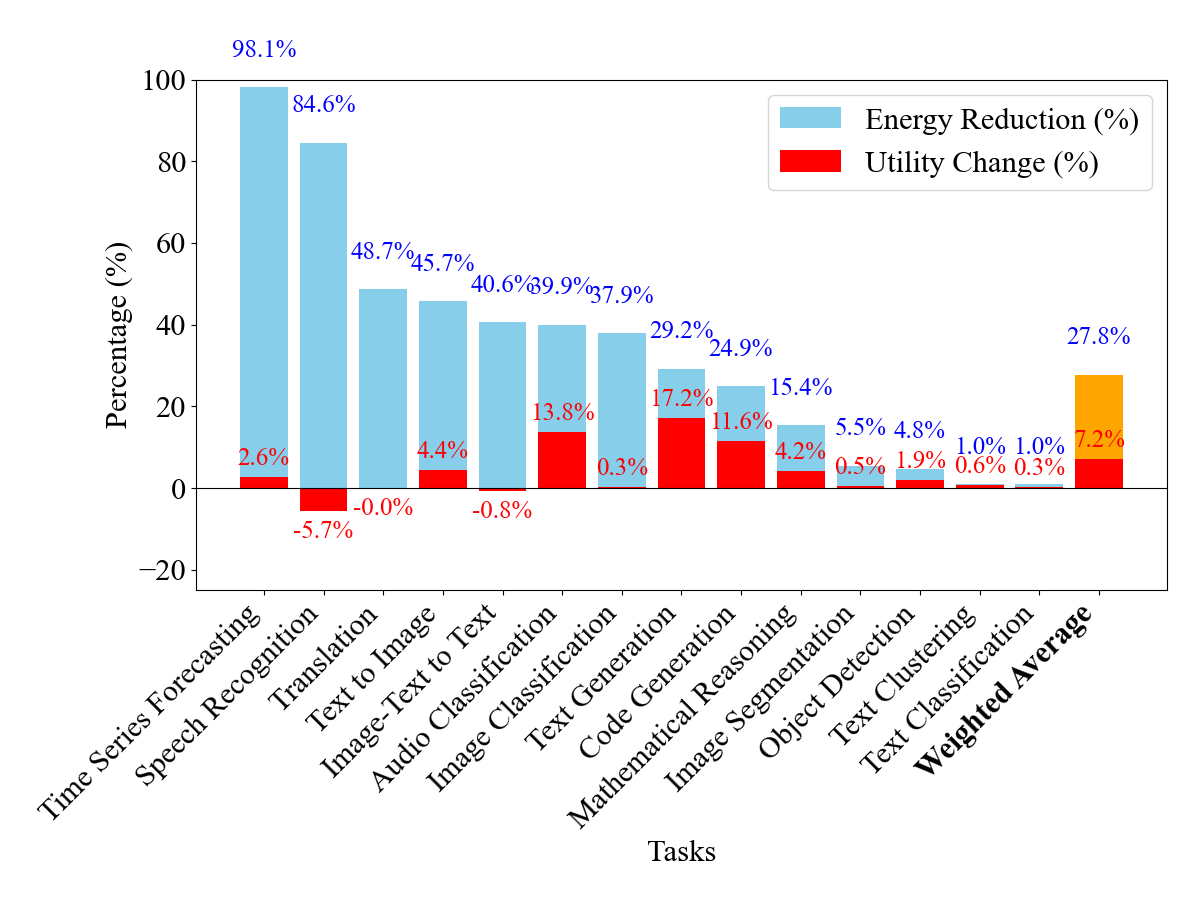}
    \caption{\textbf{Estimated Impact of Efficient Model Selection on AI Inference.} Estimated percentage reduction in energy consumption and the percentage variation in utility for major AI tasks, when applying model selection during inference. For each task, inferences with models larger than the identified \efficient model are redirected to it. The rightmost bars represent the weighted average across all considered AI tasks weighted by their usage (number of downloads).    
    \label{fig:energySobrietyPotential:scenario3}}
\end{figure}

\begin{allFigs}
    \begin{figure*}[ht!]
    \centering
    \begin{subfigure}[]{0.48\textwidth}
        \centering
        \includegraphics[width=\linewidth]{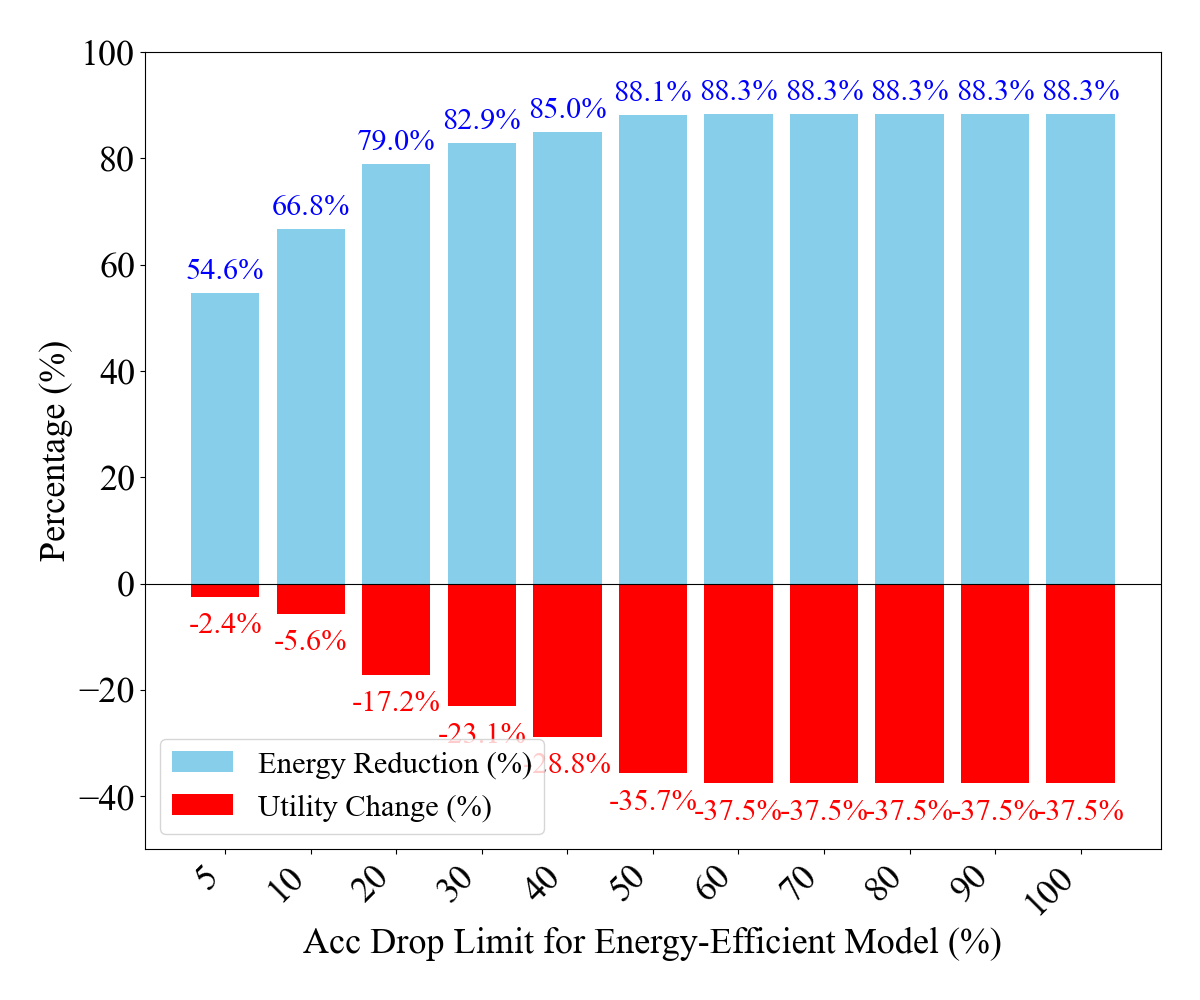}
        \caption{Key models}
        \label{fig:energySavingsAccDropScenario1}
    \end{subfigure}
    \begin{subfigure}[]{0.48\textwidth}
        \centering
        \includegraphics[width=\linewidth]{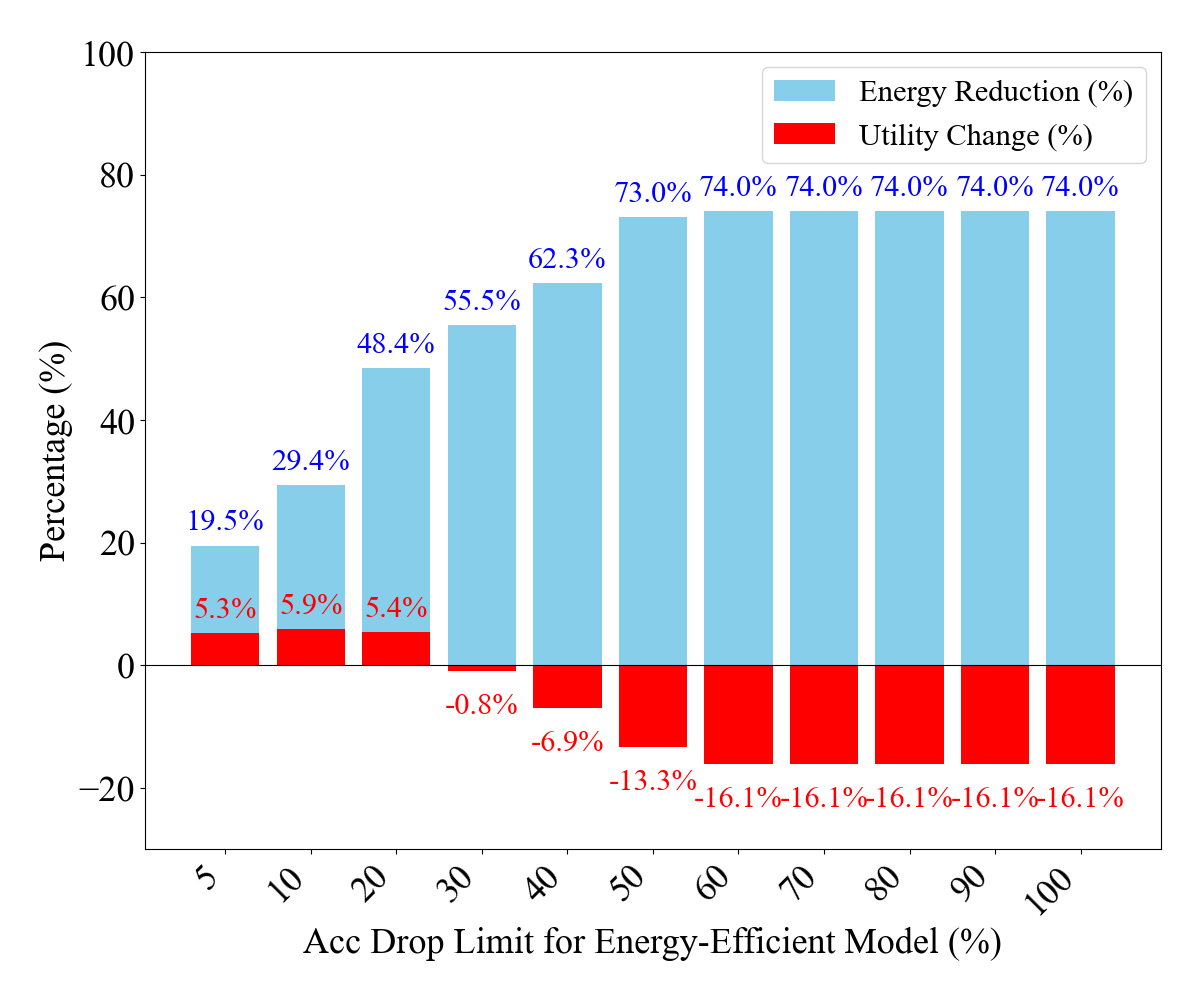}
        \caption{All models}
        \label{fig:energySavingsAccDropScenario3}
    \end{subfigure}
    \caption{\textbf{Trade-off: Energy Savings vs. Maximum Utility Drops.} Global estimated energy savings (\%) and utility variation (\%) under different maximum utility drop thresholds when choosing the \efficient model. Results are shown (a) considering only the two key models (\best and \efficient), and (b) considering all models. For each threshold, the presented value is a download-weighted average across all tasks.
    \label{fig:energySavingsAccDrop}
    }
\end{figure*}
\end{allFigs}

\textbf{Our findings indicate that model selection applied for AI inference results in a $\bm{27.8\%}$ reduction in energy consumption.} Among the evaluated tasks, two tasks present a very high potential energy saving, over than $80\%$: \textit{time series forecasting} and \textit{speech recognition}. 
For these tasks, the most commonly used models are usually large and power-hungry. Thus, the model selection and the AI inference requests redirecting towards energy-efficient models represent a substantial energy reduction (see Figure \ref{fig:TotalAIbenchmarksAnalysis}). 
Although these tasks are mature, the transition to small and efficient models has not yet been made (still in phase 3).
We also observe that other tasks
with popular large models (e.g., {\tt T5-11b} for \textit{Translation} and \textit{{\tt FLUX-dev} for Text-to-Image}) show significant energy savings, over $40\%$.


Furthermore, Figure \ref{fig:energySobrietyPotential:scenario3} presents the estimated utility variation in percentage for each task. The largest observed utility decrease was $5.7\%$ for \textit{speech recognition}. \textbf{On average, however, utility increased by $\bm{4\%}$.} Although this result may seem counter-intuitive, this is expected, since some widely used models are actually bigger and less performing than the energy-efficient model (due to reasons presented in Section~\ref{sec:model-adoption} discussing model adoption). Thus, switching to the energy-efficient model can sometimes lead to an improvement in utility. This finding aligns with previous discussions by Abio~\cite{abio2023ai} and Varoquaux et al.~\cite{varoquaux2024hype}, who challenge the "bigger-is-better" paradigm, arguing that smaller models can perform as well as, or even better than, larger models.



When evaluating energy savings under different utility drop values for choosing \efficient model, the results (see Figure \ref{fig:energySavingsAccDropScenario3}) point that the user can save up to $74.0\%$ of energy consumption applying model selection with a maximal global utility loss of $16.1\%$.


Model selection can thus be an efficient method to reduce the impact of AI, while not affecting its global utility. However, several scenarios are possible for the future. We present the estimated energy consumption for AI inference in U.S. data centers for three of them in Figure~\ref{fig:energyEstimation:Historic}. The historical data was retrieved from the U.S. Data Centers report, assuming an inference ratio of $60\%$~\cite{USEnergyDataCenter}. 
The figure presents three possible future scenarios for energy consumption: (i) a \emph{business-as-usual scenario}, reflecting the natural growth of AI and data center usage; (ii) a \emph{pessimistic scenario}, assuming widespread deployment of the \best (and often large) models due to hardware evolution; and (iii) a \emph{sobriety scenario}, where model selection redirects inference requests to the \efficient model. We considered a transition period of a year, 2025-2026, to transition to Scenarios (ii) and (iii). Thus, from 2026, we recompute the energy consumption as if all the used models were \best or \efficient.
For each scenario, we consider both an upper bound and a lower bound on the estimated energy consumption, as reported in~\cite{USEnergyDataCenter}. Our projections indicate that \textbf{using large and power-hungry models, in a pessimistic scenario, could increase energy consumption by $\bm{111.2\%}$.} On the other hand, \textbf{adopting model selection} would enhance energy sobriety, saving $27.8\%$.

For U.S. data centers, we estimate that model selection could save $\bm{16.25}$ TWh in 2025, equivalent to the annual energy production of two nuclear power reactors. \textbf{By 2028, these savings could reach $\bm{41.8}$ TWh, corresponding to the annual production of seven nuclear power reactors.}

Expanding our analysis to global data centers based on estimations from SemiAnalysis study \cite{patel2024ai}, \textbf{we project energy savings from model selection of $\bm{31.9}$ TWh in 2025 and $\bm{106}$ TWh in 2028 - equivalent to the annual production of 5 and 17 nuclear power reactors, respectively.} These findings highlight the potential of model selection for achieving energy sobriety during AI inference at scale.

\begin{figure}[ht]
    \centering
    \includegraphics[width=\linewidth]{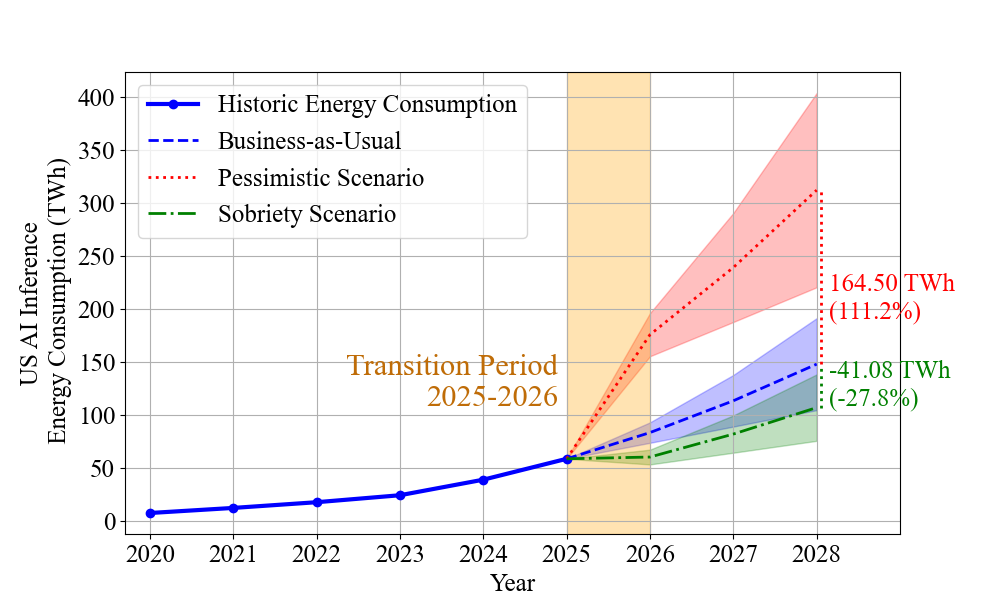}
    \caption{\textbf{Projected US Data Center Energy Consumption for AI Inference Under 3 Different Scenarios.} US data centers AI inference energy projection over time for three different scenarios: (i) business as usual, (ii) sobriety using model selection, (iii) use of \best model with the needed hardware. The values for (i) were extracted from \cite{USEnergyDataCenter}. We considered a one-year transition to reach scenario (ii), applying model selection globally, or (iii), using the \best model, (during the year 2025). 
    \label{fig:energyEstimation:Historic}
    }
\end{figure}

\begin{alternativeFigs}
    \begin{figure}
    \centering
    \begin{subfigure}[]{0.47\textwidth}
    \centering
        \includegraphics[width=\linewidth]{Figs/HF_Acc_Eff_sobriety_estimation_-25_100_0.1_scenario3.png}
        \caption{Estimated energy savings and utility variation, in percentage, for each task when adopting model selection. \tiago{The \efficient model was chosen with a maximal utility drop of $10\%$}. The weighted average uses the number of downloads of each task to weight the values.}
        \label{fig:energySobrietyPotential:scenario3}
    \end{subfigure}
    \hfill
    \begin{subfigure}[]{0.47\textwidth}
    \centering
    \includegraphics[width=\linewidth]{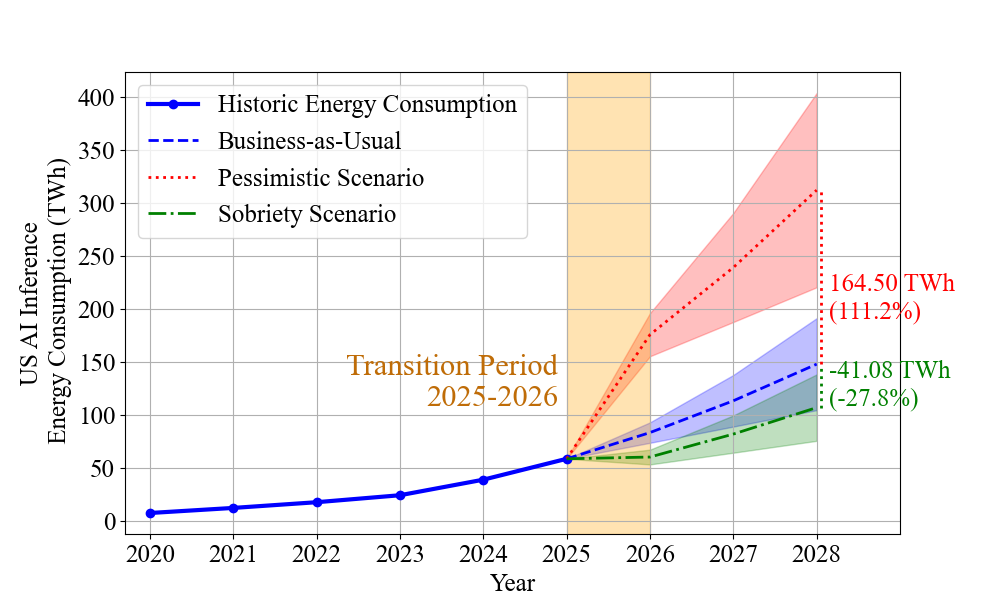}
    \caption{US data centres AI inference energy over time: history and projection for three different scenarios: (i) business as usual, (ii) sobriety using model selection, (iii) \best model with the needed hardware. The historical data was extracted from \cite{USEnergyDataCenter}. 
    \label{fig:energyEstimation:Historic} }
    \end{subfigure}
    \caption{\tiago{2nd Option:} Energy sobriety estimation through model selection on inference for several tasks. \fred{captions have to be improvecd}}
    \label{fig:energySobrietyPotential}
\end{figure}
\end{alternativeFigs}

\section{Conclusion}
\label{sec:conclusion}

Given the widespread adoption of AI and the significant energy consumption of many popular models, in this work, we investigated how model selection can contribute to energy sobriety. Specifically, model selection aims to identify energy-efficient models that (i) have a small size, thus reducing energy consumption, and (ii) maintain high utility, thus ensuring good performance.

To achieve this, we first analyzed model collection platforms such as \textit{Hugging Face} and \textit{Papers with Code} to identify the most commonly used models in the developer community. We then analyzed the utility-size trade-off across several tasks and observed a diminishing marginal gain that can be exploited by model selection to reduce overall energy consumption.

Based on this pattern, we estimated the potential energy gain from selecting energy-efficient models for inference tasks. Our estimates suggest a reduction in energy consumption of $27.8\%$. In the United States, this reduction is equivalent to the annual production of two nuclear power reactors.

As the urgency of climate change grows, the adoption of sustainable AI practices is critical. We believe that the responsible and efficient use of AI plays a key role in mitigating environmental impact. As a future direction, we aim to expand our analysis to consider the entire life cycle of AI models, from data collection to deployment.

\bibliographystyle{IEEEtran}

\begin{thebibliography}{10}
\providecommand{\url}[1]{#1}
\csname url@samestyle\endcsname
\providecommand{\newblock}{\relax}
\providecommand{\bibinfo}[2]{#2}
\providecommand{\BIBentrySTDinterwordspacing}{\spaceskip=0pt\relax}
\providecommand{\BIBentryALTinterwordstretchfactor}{4}
\providecommand{\BIBentryALTinterwordspacing}{\spaceskip=\fontdimen2\font plus
\BIBentryALTinterwordstretchfactor\fontdimen3\font minus
  \fontdimen4\font\relax}
\providecommand{\BIBforeignlanguage}[2]{{%
\expandafter\ifx\csname l@#1\endcsname\relax
\typeout{** WARNING: IEEEtran.bst: No hyphenation pattern has been}%
\typeout{** loaded for the language `#1'. Using the pattern for}%
\typeout{** the default language instead.}%
\else
\language=\csname l@#1\endcsname
\fi
#2}}
\providecommand{\BIBdecl}{\relax}
\BIBdecl

\bibitem{EpochNotableModels2024}
\BIBentryALTinterwordspacing
{Epoch AI}, ``Data on notable ai models,'' 2024, accessed: 2025-01-16.
  [Online]. Available: \url{https://epoch.ai/data/notable-ai-models}
\BIBentrySTDinterwordspacing

\bibitem{wu2022sustainable}
C.-J. Wu, R.~Raghavendra, U.~Gupta, B.~Acun, N.~Ardalani, K.~Maeng, G.~Chang,
  F.~Aga, J.~Huang, C.~Bai \emph{et~al.}, ``Sustainable ai: Environmental
  implications, challenges and opportunities,'' \emph{Proceedings of Machine
  Learning and Systems}, vol.~4, pp. 795--813, 2022.

\bibitem{USEnergyDataCenter}
\BIBentryALTinterwordspacing
A.~Shehabi, S.~J. Smith, A.~Hubbard, A.~Newkirk, N.~Lei, M.~A. Siddik,
  B.~Holecek, J.~G. Koomey, E.~R. Masanet, and D.~A. Sartor, ``2024 united
  states data center energy usage report,'' 19/12/2024 2024. [Online].
  Available:
  \url{https://eta-publications.lbl.gov/sites/default/files/2024-12/lbnl-2024-united-states-data-center-energy-usage-report.pdf}
\BIBentrySTDinterwordspacing

\bibitem{paris2015paris}
A.~Paris, ``Paris agreement to the united nations framework convention o n
  climate change,'' \emph{Adopted Dec}, vol.~12, 2015.

\bibitem{lee2023ipcc}
H.~Lee, K.~Calvin, D.~Dasgupta, G.~Krinner, A.~Mukherji, P.~Thorne, C.~Trisos,
  J.~Romero, P.~Aldunce, K.~Barret \emph{et~al.}, ``Ipcc, 2023: Climate change
  2023: Synthesis report, summary for policymakers. contribution of working
  groups i, ii and iii to the sixth assessment report of the intergovernmental
  panel on climate change [core writing team, h. lee and j. romero (eds.)].
  ipcc, geneva, switzerland.'' 2023.

\bibitem{tao2020challenges}
Y.~Tao, R.~Ma, M.-L. Shyu, and S.-C. Chen, ``Challenges in energy-efficient
  deep neural network training with fpga,'' in \emph{Proceedings of the
  IEEE/CVF conference on computer vision and pattern recognition workshops},
  2020, pp. 400--401.

\bibitem{zhang2018exploring}
B.~Zhang, A.~Davoodi, and Y.~H. Hu, ``Exploring energy and accuracy tradeoff in
  structure simplification of trained deep neural networks,'' \emph{IEEE
  Journal on Emerging and Selected Topics in Circuits and Systems}, vol.~8,
  no.~4, pp. 836--848, 2018.

\bibitem{mehta2019espnetv2}
S.~Mehta, M.~Rastegari, L.~Shapiro, and H.~Hajishirzi, ``Espnetv2: A
  light-weight, power efficient, and general purpose convolutional neural
  network,'' in \emph{Proceedings of the IEEE/CVF conference on computer vision
  and pattern recognition}, 2019, pp. 9190--9200.

\bibitem{asperti2021survey}
A.~Asperti, D.~Evangelista, and E.~Loli~Piccolomini, ``A survey on variational
  autoencoders from a green ai perspective,'' \emph{SN Computer Science},
  vol.~2, no.~4, p. 301, 2021.

\bibitem{yu2022energy}
J.-R. Yu, C.-H. Chen, T.-W. Huang, J.-J. Lu, C.-R. Chung, T.-W. Lin, M.-H. Wu,
  Y.-J. Tseng, and H.-Y. Wang, ``Energy efficiency of inference algorithms for
  clinical laboratory data sets: Green artificial intelligence study,''
  \emph{Journal of Medical Internet Research}, vol.~24, no.~1, p. e28036, 2022.

\bibitem{patterson2022carbon}
D.~Patterson, J.~Gonzalez, U.~H{\"o}lzle, Q.~Le, C.~Liang, L.-M. Munguia,
  D.~Rothchild, D.~R. So, M.~Texier, and J.~Dean, ``The carbon footprint of
  machine learning training will plateau, then shrink,'' \emph{Computer},
  vol.~55, no.~7, pp. 18--28, 2022.

\bibitem{deepseek}
``Deepseek,'' \url{https://www.deepseek.com/}.

\bibitem{huggingface}
``Hugging face,'' \url{https://huggingface.co/}.

\bibitem{paperswithcode}
``Papers with code: The latest in machine learning,''
  \url{https://paperswithcode.com/}, accessed: 2024-08-06.

\bibitem{krizhevsky2012imagenet}
A.~Krizhevsky, I.~Sutskever, and G.~E. Hinton, ``Imagenet classification with
  deep convolutional neural networks,'' \emph{Advances in neural information
  processing systems}, vol.~25, 2012.

\bibitem{turgot}
G.~Faccarello, ``Anne-robert-jacques turgot (1727--1781),'' \emph{Handbook on
  the history of economic analysis}, vol.~1, pp. 73--82, 2016.

\bibitem{frantar2023sparsegpt}
E.~Frantar and D.~Alistarh, ``Sparsegpt: Massive language models can be
  accurately pruned in one-shot,'' in \emph{International Conference on Machine
  Learning}.\hskip 1em plus 0.5em minus 0.4em\relax PMLR, 2023, pp.
  10\,323--10\,337.

\bibitem{shao2023omniquant}
W.~Shao, M.~Chen, Z.~Zhang, P.~Xu, L.~Zhao, Z.~Li, K.~Zhang, P.~Gao, Y.~Qiao,
  and P.~Luo, ``Omniquant: Omnidirectionally calibrated quantization for large
  language models,'' \emph{arXiv preprint arXiv:2308.13137}, 2023.

\bibitem{ma2023llm}
X.~Ma, G.~Fang, and X.~Wang, ``Llm-pruner: On the structural pruning of large
  language models,'' \emph{Advances in neural information processing systems},
  vol.~36, pp. 21\,702--21\,720, 2023.

\bibitem{perrault2024artificial}
R.~Perrault and J.~Clark, ``Artificial intelligence index report 2024,'' 2024.

\bibitem{bommasani2021opportunities}
R.~Bommasani, D.~A. Hudson, E.~Adeli, R.~Altman, S.~Arora, S.~von Arx, M.~S.
  Bernstein, J.~Bohg, A.~Bosselut, E.~Brunskill \emph{et~al.}, ``On the
  opportunities and risks of foundation models,'' \emph{arXiv preprint
  arXiv:2108.07258}, 2021.

\bibitem{open-llm-leaderboard-v1}
E.~Beeching, C.~Fourrier, N.~Habib, S.~Han, N.~Lambert, N.~Rajani,
  O.~Sanseviero, L.~Tunstall, and T.~Wolf, ``Open llm leaderboard
  (2023-2024),''
  \url{https://huggingface.co/spaces/open-llm-leaderboard-old/open_llm_leaderboard},
  2023.

\bibitem{zheng2024judging}
L.~Zheng, W.-L. Chiang, Y.~Sheng, S.~Zhuang, Z.~Wu, Y.~Zhuang, Z.~Lin, Z.~Li,
  D.~Li, E.~Xing \emph{et~al.}, ``Judging llm-as-a-judge with mt-bench and
  chatbot arena,'' \emph{Advances in Neural Information Processing Systems},
  vol.~36, 2024.

\bibitem{fan2023nphardeval}
L.~Fan, W.~Hua, L.~Li, H.~Ling, Y.~Zhang, and L.~Hemphill, ``Nphardeval:
  Dynamic benchmark on reasoning ability of large language models via
  complexity classes,'' 2023.

\bibitem{bigcode-evaluation-harness}
L.~Ben~Allal, N.~Muennighoff, L.~Kumar~Umapathi, B.~Lipkin, and L.~von Werra,
  ``A framework for the evaluation of code generation models,''
  \url{https://github.com/bigcode-project/bigcode-evaluation-harness}, 2022.

\bibitem{muennighoff2022mteb}
\BIBentryALTinterwordspacing
N.~Muennighoff, N.~Tazi, L.~Magne, and N.~Reimers, ``Mteb: Massive text
  embedding benchmark,'' \emph{arXiv preprint arXiv:2210.07316}, 2022.
  [Online]. Available: \url{https://arxiv.org/abs/2210.07316}
\BIBentrySTDinterwordspacing

\bibitem{bojar-etal-2014-findings}
\BIBentryALTinterwordspacing
O.~Bojar, C.~Buck, C.~Federmann, B.~Haddow, P.~Koehn, J.~Leveling, C.~Monz,
  P.~Pecina, M.~Post, H.~Saint-Amand, R.~Soricut, L.~Specia, and A.~Tamchyna,
  ``Findings of the 2014 workshop on statistical machine translation,'' in
  \emph{Proceedings of the Ninth Workshop on Statistical Machine
  Translation}.\hskip 1em plus 0.5em minus 0.4em\relax Baltimore, Maryland,
  USA: Association for Computational Linguistics, Jun. 2014, pp. 12--58.
  [Online]. Available: \url{https://aclanthology.org/W14-3302}
\BIBentrySTDinterwordspacing

\bibitem{open-od-leaderboard}
A.~R. Rafael~Padilla and the Hugging Face~Team, ``Open object detection
  leaderboard,''
  \url{https://huggingface.co/spaces/rafaelpadilla/object_detection_leaderboard},
  2023.

\bibitem{russakovsky2015imagenet}
O.~Russakovsky, J.~Deng, H.~Su, J.~Krause, S.~Satheesh, S.~Ma, Z.~Huang,
  A.~Karpathy, A.~Khosla, M.~Bernstein \emph{et~al.}, ``Imagenet large scale
  visual recognition challenge,'' \emph{International journal of computer
  vision}, vol. 115, pp. 211--252, 2015.

\bibitem{Zhou_2017_CVPR}
B.~Zhou, H.~Zhao, X.~Puig, S.~Fidler, A.~Barriuso, and A.~Torralba, ``Scene
  parsing through ade20k dataset,'' in \emph{Proceedings of the IEEE Conference
  on Computer Vision and Pattern Recognition (CVPR)}, July 2017.

\bibitem{open-asr-leaderboard}
V.~Srivastav, S.~Majumdar, N.~Koluguri, A.~Moumen, S.~Gandhi \emph{et~al.},
  ``Open automatic speech recognition leaderboard,''
  \url{https://huggingface.co/spaces/hf-audio/open_asr_leaderboard}, 2023.

\bibitem{ARCH}
M.~La~Quatra, A.~Koudounas, L.~Vaiani, E.~Baralis, L.~Cagliero, P.~Garza, and
  S.~M. Siniscalchi, ``Benchmarking representations for speech, music, and
  acoustic events,'' in \emph{2024 IEEE International Conference on Acoustics,
  Speech, and Signal Processing Workshops (ICASSPW)}, 2024, pp. 505--509.

\bibitem{ku2024imagenhub}
\BIBentryALTinterwordspacing
M.~Ku, T.~Li, K.~Zhang, Y.~Lu, X.~Fu, W.~Zhuang, and W.~Chen, ``Imagenhub:
  Standardizing the evaluation of conditional image generation models,'' in
  \emph{The Twelfth International Conference on Learning Representations},
  2024. [Online]. Available: \url{https://openreview.net/forum?id=OuV9ZrkQlc}
\BIBentrySTDinterwordspacing

\bibitem{yue2023mmmu}
X.~Yue, Y.~Ni, K.~Zhang, T.~Zheng, R.~Liu, G.~Zhang, S.~Stevens, D.~Jiang,
  W.~Ren, Y.~Sun, C.~Wei, B.~Yu, R.~Yuan, R.~Sun, M.~Yin, B.~Zheng, Z.~Yang,
  Y.~Liu, W.~Huang, H.~Sun, Y.~Su, and W.~Chen, ``Mmmu: A massive
  multi-discipline multimodal understanding and reasoning benchmark for expert
  {AGI},'' in \emph{Proceedings of the IEEE/CVF Conference on Computer Vision
  and Pattern Recognition (CVPR)}, 2024.

\bibitem{zhou2021informer}
H.~Zhou, S.~Zhang, J.~Peng, S.~Zhang, J.~Li, H.~Xiong, and W.~Zhang,
  ``Informer: Beyond efficient transformer for long sequence time-series
  forecasting,'' in \emph{Proceedings of the AAAI conference on artificial
  intelligence}, vol.~35, no.~12, 2021, pp. 11\,106--11\,115.

\bibitem{kaplan2020scaling}
J.~Kaplan, S.~McCandlish, T.~Henighan, T.~B. Brown, B.~Chess, R.~Child,
  S.~Gray, A.~Radford, J.~Wu, and D.~Amodei, ``Scaling laws for neural language
  models,'' \emph{arXiv preprint arXiv:2001.08361}, 2020.

\bibitem{desislavov2023trends}
R.~Desislavov, F.~Mart{\'\i}nez-Plumed, and J.~Hern{\'a}ndez-Orallo, ``Trends
  in ai inference energy consumption: Beyond the performance-vs-parameter laws
  of deep learning,'' \emph{Sustainable Computing: Informatics and Systems},
  vol.~38, p. 100857, 2023.

\bibitem{labrecque2017habit}
J.~S. Labrecque, W.~Wood, D.~T. Neal, and N.~Harrington, ``Habit slips: When
  consumers unintentionally resist new products,'' \emph{Journal of the Academy
  of Marketing Science}, vol.~45, pp. 119--133, 2017.

\bibitem{heidenreich2016innovations}
S.~Heidenreich and T.~Kraemer, ``Innovations—doomed to fail? investigating
  strategies to overcome passive innovation resistance,'' \emph{Journal of
  Product Innovation Management}, vol.~33, no.~3, pp. 277--297, 2016.

\bibitem{ram1989consumer}
S.~Ram and J.~N. Sheth, ``Consumer resistance to innovations: the marketing
  problem and its solutions,'' \emph{Journal of consumer marketing}, vol.~6,
  no.~2, pp. 5--14, 1989.

\bibitem{similarweb2025}
\BIBentryALTinterwordspacing
{SimilarWeb}, ``Website traffic analysis,'' 2025, accessed: 2025-03-20.
  [Online]. Available: \url{https://www.similarweb.com}
\BIBentrySTDinterwordspacing

\bibitem{luccioni2023estimating}
A.~S. Luccioni, S.~Viguier, and A.-L. Ligozat, ``Estimating the carbon
  footprint of bloom, a 176b parameter language model,'' \emph{Journal of
  Machine Learning Research}, vol.~24, no. 253, pp. 1--15, 2023.

\bibitem{luccioni2024power}
S.~Luccioni, Y.~Jernite, and E.~Strubell, ``Power hungry processing: Watts
  driving the cost of ai deployment?'' in \emph{The 2024 ACM Conference on
  Fairness, Accountability, and Transparency}, 2024, pp. 85--99.

\bibitem{yang2024harnessing}
J.~Yang, H.~Jin, R.~Tang, X.~Han, Q.~Feng, H.~Jiang, S.~Zhong, B.~Yin, and
  X.~Hu, ``Harnessing the power of llms in practice: A survey on chatgpt and
  beyond,'' \emph{ACM Transactions on Knowledge Discovery from Data}, vol.~18,
  no.~6, pp. 1--32, 2024.

\bibitem{dodge2022measuring}
J.~Dodge, T.~Prewitt, R.~Tachet~des Combes, E.~Odmark, R.~Schwartz,
  E.~Strubell, A.~S. Luccioni, N.~A. Smith, N.~DeCario, and W.~Buchanan,
  ``Measuring the carbon intensity of ai in cloud instances,'' in
  \emph{Proceedings of the 2022 ACM conference on fairness, accountability, and
  transparency}, 2022, pp. 1877--1894.

\bibitem{anthony2020carbontracker}
L.~F.~W. Anthony, B.~Kanding, and R.~Selvan, ``Carbontracker: Tracking and
  predicting the carbon footprint of training deep learning models,'' ICML
  Workshop on Challenges in Deploying and monitoring Machine Learning Systems,
  July 2020, arXiv:2007.03051.

\bibitem{benoit_courty_2024_11171501}
\BIBentryALTinterwordspacing
B.~Courty, V.~Schmidt, S.~Luccioni, Goyal-Kamal, MarionCoutarel, B.~Feld,
  J.~Lecourt, LiamConnell, A.~Saboni, Inimaz, supatomic, M.~Léval, L.~Blanche,
  A.~Cruveiller, ouminasara, F.~Zhao, A.~Joshi, A.~Bogroff, H.~de~Lavoreille,
  N.~Laskaris, E.~Abati, D.~Blank, Z.~Wang, A.~Catovic, M.~Alencon, M.~Stechly,
  C.~Bauer, L.~O.~N. de~Araújo, JPW, and MinervaBooks, ``mlco2/codecarbon:
  v2.4.1,'' May 2024. [Online]. Available:
  \url{https://doi.org/10.5281/zenodo.11171501}
\BIBentrySTDinterwordspacing

\bibitem{zeus:nsdi23}
J.~You, J.-W. Chung, and M.~Chowdhury, ``Zeus: Understanding and optimizing
  {GPU} energy consumption of {DNN} training,'' in \emph{USENIX NSDI}, 2023.

\bibitem{scaphandre}
\BIBentryALTinterwordspacing
B.~Petit, ``scaphandre,'' 2023. [Online]. Available:
  \url{https://github.com/hubblo-org/scaphandre}
\BIBentrySTDinterwordspacing

\bibitem{powerspy2}
\BIBentryALTinterwordspacing
ALCIOM, ``Powerspy2: An advanced power analyzer,'' n.d., accessed: 2025-03-10.
  [Online]. Available:
  \url{https://www.alciom.com/en/our-trades/products/powerspy2/}
\BIBentrySTDinterwordspacing

\bibitem{abio2023ai}
B.~Abio, ``In ai, is bigger better?'' \emph{Nature}, vol. 615, no. 7951, pp.
  202--205, 2023.

\bibitem{varoquaux2024hype}
G.~Varoquaux, A.~S. Luccioni, and M.~Whittaker, ``Hype, sustainability, and the
  price of the bigger-is-better paradigm in ai,'' \emph{arXiv preprint
  arXiv:2409.14160}, 2024.

\bibitem{patel2024ai}
D.~Patel, D.~Nishball, and J.~E. Ontiveros, ``Ai datacenter energy dilemma-race
  for ai datacenter space,'' \emph{SemiAnalysis, March}, vol.~13, 2024.

\end{thebibliography}

\begin{thebibliography}{10}
\providecommand{\url}[1]{#1}
\csname url@samestyle\endcsname
\providecommand{\newblock}{\relax}
\providecommand{\bibinfo}[2]{#2}
\providecommand{\BIBentrySTDinterwordspacing}{\spaceskip=0pt\relax}
\providecommand{\BIBentryALTinterwordstretchfactor}{4}
\providecommand{\BIBentryALTinterwordspacing}{\spaceskip=\fontdimen2\font plus
\BIBentryALTinterwordstretchfactor\fontdimen3\font minus
  \fontdimen4\font\relax}
\providecommand{\BIBforeignlanguage}[2]{{%
\expandafter\ifx\csname l@#1\endcsname\relax
\typeout{** WARNING: IEEEtran.bst: No hyphenation pattern has been}%
\typeout{** loaded for the language `#1'. Using the pattern for}%
\typeout{** the default language instead.}%
\else
\language=\csname l@#1\endcsname
\fi
#2}}
\providecommand{\BIBdecl}{\relax}
\BIBdecl

\bibitem{anthony2020carbontracker}
L.~F.~W. Anthony, B.~Kanding, and R.~Selvan, ``Carbontracker: Tracking and
  predicting the carbon footprint of training deep learning models,'' ICML
  Workshop on Challenges in Deploying and monitoring Machine Learning Systems,
  July 2020, arXiv:2007.03051.

\bibitem{nvidia2022nvidia}
S.~NVIDIA, ``Nvidia management library (nvml),'' 2022.

\bibitem{khan2018rapl}
K.~N. Khan, M.~Hirki, T.~Niemi, J.~K. Nurminen, and Z.~Ou, ``Rapl in action:
  Experiences in using rapl for power measurements,'' \emph{ACM Transactions on
  Modeling and Performance Evaluation of Computing Systems (TOMPECS)}, vol.~3,
  no.~2, pp. 1--26, 2018.

\bibitem{geissler2024power}
D.~Gei{\ss}ler, B.~Zhou, M.~Liu, S.~Suh, and P.~Lukowicz, ``The power of
  training: How different neural network setups influence the energy demand,''
  in \emph{International Conference on Architecture of Computing
  Systems}.\hskip 1em plus 0.5em minus 0.4em\relax Springer, 2024, pp. 33--47.

\bibitem{douwes2021energy}
C.~Douwes, P.~Esling, and J.-P. Briot, ``Energy consumption of deep generative
  audio models,'' \emph{arXiv preprint arXiv:2107.02621}, 2021.

\bibitem{bannour2021evaluating}
N.~Bannour, S.~Ghannay, A.~N{\'e}v{\'e}ol, and A.-L. Ligozat, ``Evaluating the
  carbon footprint of nlp methods: a survey and analysis of existing tools,''
  in \emph{Proceedings of the second workshop on simple and efficient natural
  language processing}, 2021, pp. 11--21.

\bibitem{rodriguez2024evaluating}
C.~Rodriguez, L.~Degioanni, L.~Kameni, R.~Vidal, and G.~Neglia, ``Evaluating
  the energy consumption of machine learning: Systematic literature review and
  experiments,'' \emph{arXiv preprint arXiv:2408.15128}, 2024.

\bibitem{powerspy2}
\BIBentryALTinterwordspacing
ALCIOM, ``Powerspy2: An advanced power analyzer,'' n.d., accessed: 2025-03-10.
  [Online]. Available:
  \url{https://www.alciom.com/en/our-trades/products/powerspy2/}
\BIBentrySTDinterwordspacing

\bibitem{ecologits}
G.~Impact, ``Ecologits: A tool for measuring energy consumption of machine
  learning models,'' \url{https://github.com/genai-impact/ecologits}, 2024,
  accessed: 2024-11-27.

\bibitem{similarweb2025}
\BIBentryALTinterwordspacing
{SimilarWeb}, ``Website traffic analysis,'' 2025, accessed: 2025-03-20.
  [Online]. Available: \url{https://www.similarweb.com}
\BIBentrySTDinterwordspacing

\bibitem{raffel2020exploring}
C.~Raffel, N.~Shazeer, A.~Roberts, K.~Lee, S.~Narang, M.~Matena, Y.~Zhou,
  W.~Li, and P.~J. Liu, ``Exploring the limits of transfer learning with a
  unified text-to-text transformer,'' \emph{Journal of machine learning
  research}, vol.~21, no. 140, pp. 1--67, 2020.

\bibitem{perrault2024artificial}
R.~Perrault and J.~Clark, ``Artificial intelligence index report 2024,'' 2024.

\bibitem{open-llm-leaderboard-v1}
E.~Beeching, C.~Fourrier, N.~Habib, S.~Han, N.~Lambert, N.~Rajani,
  O.~Sanseviero, L.~Tunstall, and T.~Wolf, ``Open llm leaderboard
  (2023-2024),''
  \url{https://huggingface.co/spaces/open-llm-leaderboard-old/open_llm_leaderboard},
  2023.

\bibitem{eval-harness}
\BIBentryALTinterwordspacing
L.~Gao, J.~Tow, S.~Biderman, S.~Black, A.~DiPofi, C.~Foster, L.~Golding,
  J.~Hsu, K.~McDonell, N.~Muennighoff, J.~Phang, L.~Reynolds, E.~Tang,
  A.~Thite, B.~Wang, K.~Wang, and A.~Zou, ``A framework for few-shot language
  model evaluation,'' Sep. 2021. [Online]. Available:
  \url{https://doi.org/10.5281/zenodo.5371628}
\BIBentrySTDinterwordspacing

\bibitem{chiang2024chatbot}
W.-L. Chiang, L.~Zheng, Y.~Sheng, A.~N. Angelopoulos, T.~Li, D.~Li, H.~Zhang,
  B.~Zhu, M.~Jordan, J.~E. Gonzalez \emph{et~al.}, ``Chatbot arena: An open
  platform for evaluating llms by human preference,'' \emph{arXiv preprint
  arXiv:2403.04132}, 2024.

\bibitem{zheng2024judging}
L.~Zheng, W.-L. Chiang, Y.~Sheng, S.~Zhuang, Z.~Wu, Y.~Zhuang, Z.~Lin, Z.~Li,
  D.~Li, E.~Xing \emph{et~al.}, ``Judging llm-as-a-judge with mt-bench and
  chatbot arena,'' \emph{Advances in Neural Information Processing Systems},
  vol.~36, 2024.

\bibitem{fan2023nphardeval}
L.~Fan, W.~Hua, L.~Li, H.~Ling, Y.~Zhang, and L.~Hemphill, ``Nphardeval:
  Dynamic benchmark on reasoning ability of large language models via
  complexity classes,'' 2023.

\bibitem{bigcode-evaluation-harness}
L.~Ben~Allal, N.~Muennighoff, L.~Kumar~Umapathi, B.~Lipkin, and L.~von Werra,
  ``A framework for the evaluation of code generation models,''
  \url{https://github.com/bigcode-project/bigcode-evaluation-harness}, 2022.

\bibitem{bojar-etal-2014-findings}
\BIBentryALTinterwordspacing
O.~Bojar, C.~Buck, C.~Federmann, B.~Haddow, P.~Koehn, J.~Leveling, C.~Monz,
  P.~Pecina, M.~Post, H.~Saint-Amand, R.~Soricut, L.~Specia, and A.~Tamchyna,
  ``Findings of the 2014 workshop on statistical machine translation,'' in
  \emph{Proceedings of the Ninth Workshop on Statistical Machine
  Translation}.\hskip 1em plus 0.5em minus 0.4em\relax Baltimore, Maryland,
  USA: Association for Computational Linguistics, Jun. 2014, pp. 12--58.
  [Online]. Available: \url{https://aclanthology.org/W14-3302}
\BIBentrySTDinterwordspacing

\bibitem{muennighoff2022mteb}
\BIBentryALTinterwordspacing
N.~Muennighoff, N.~Tazi, L.~Magne, and N.~Reimers, ``Mteb: Massive text
  embedding benchmark,'' \emph{arXiv preprint arXiv:2210.07316}, 2022.
  [Online]. Available: \url{https://arxiv.org/abs/2210.07316}
\BIBentrySTDinterwordspacing

\bibitem{open-od-leaderboard}
A.~R. Rafael~Padilla and the Hugging Face~Team, ``Open object detection
  leaderboard,''
  \url{https://huggingface.co/spaces/rafaelpadilla/object_detection_leaderboard},
  2023.

\bibitem{russakovsky2015imagenet}
O.~Russakovsky, J.~Deng, H.~Su, J.~Krause, S.~Satheesh, S.~Ma, Z.~Huang,
  A.~Karpathy, A.~Khosla, M.~Bernstein \emph{et~al.}, ``Imagenet large scale
  visual recognition challenge,'' \emph{International journal of computer
  vision}, vol. 115, pp. 211--252, 2015.

\bibitem{rw2019timm}
R.~Wightman, ``Pytorch image models,''
  \url{https://github.com/rwightman/pytorch-image-models}, 2019.

\bibitem{Zhou_2017_CVPR}
B.~Zhou, H.~Zhao, X.~Puig, S.~Fidler, A.~Barriuso, and A.~Torralba, ``Scene
  parsing through ade20k dataset,'' in \emph{Proceedings of the IEEE Conference
  on Computer Vision and Pattern Recognition (CVPR)}, July 2017.

\bibitem{ku2024imagenhub}
\BIBentryALTinterwordspacing
M.~Ku, T.~Li, K.~Zhang, Y.~Lu, X.~Fu, W.~Zhuang, and W.~Chen, ``Imagenhub:
  Standardizing the evaluation of conditional image generation models,'' in
  \emph{The Twelfth International Conference on Learning Representations},
  2024. [Online]. Available: \url{https://openreview.net/forum?id=OuV9ZrkQlc}
\BIBentrySTDinterwordspacing

\bibitem{open-asr-leaderboard}
V.~Srivastav, S.~Majumdar, N.~Koluguri, A.~Moumen, S.~Gandhi \emph{et~al.},
  ``Open automatic speech recognition leaderboard,''
  \url{https://huggingface.co/spaces/hf-audio/open_asr_leaderboard}, 2023.

\bibitem{gandhi2022esbbenchmarkmultidomainendtoend}
\BIBentryALTinterwordspacing
S.~Gandhi, P.~von Platen, and A.~M. Rush, ``Esb: A benchmark for multi-domain
  end-to-end speech recognition,'' 2022. [Online]. Available:
  \url{https://arxiv.org/abs/2210.13352}
\BIBentrySTDinterwordspacing

\bibitem{ARCH}
M.~La~Quatra, A.~Koudounas, L.~Vaiani, E.~Baralis, L.~Cagliero, P.~Garza, and
  S.~M. Siniscalchi, ``Benchmarking representations for speech, music, and
  acoustic events,'' in \emph{2024 IEEE International Conference on Acoustics,
  Speech, and Signal Processing Workshops (ICASSPW)}, 2024, pp. 505--509.

\bibitem{yue2023mmmu}
X.~Yue, Y.~Ni, K.~Zhang, T.~Zheng, R.~Liu, G.~Zhang, S.~Stevens, D.~Jiang,
  W.~Ren, Y.~Sun, C.~Wei, B.~Yu, R.~Yuan, R.~Sun, M.~Yin, B.~Zheng, Z.~Yang,
  Y.~Liu, W.~Huang, H.~Sun, Y.~Su, and W.~Chen, ``Mmmu: A massive
  multi-discipline multimodal understanding and reasoning benchmark for expert
  {AGI},'' in \emph{Proceedings of the IEEE/CVF Conference on Computer Vision
  and Pattern Recognition (CVPR)}, 2024.

\bibitem{gemmeke2017audio}
J.~F. Gemmeke, D.~P. Ellis, D.~Freedman, A.~Jansen, W.~Lawrence, R.~C. Moore,
  M.~Plakal, and M.~Ritter, ``Audio set: An ontology and human-labeled dataset
  for audio events,'' in \emph{2017 IEEE international conference on acoustics,
  speech and signal processing (ICASSP)}.\hskip 1em plus 0.5em minus
  0.4em\relax IEEE, 2017, pp. 776--780.

\end{thebibliography}


\begin{IEEEbiography}[{\includegraphics[width=1in,height=1.25in,clip,keepaspectratio]{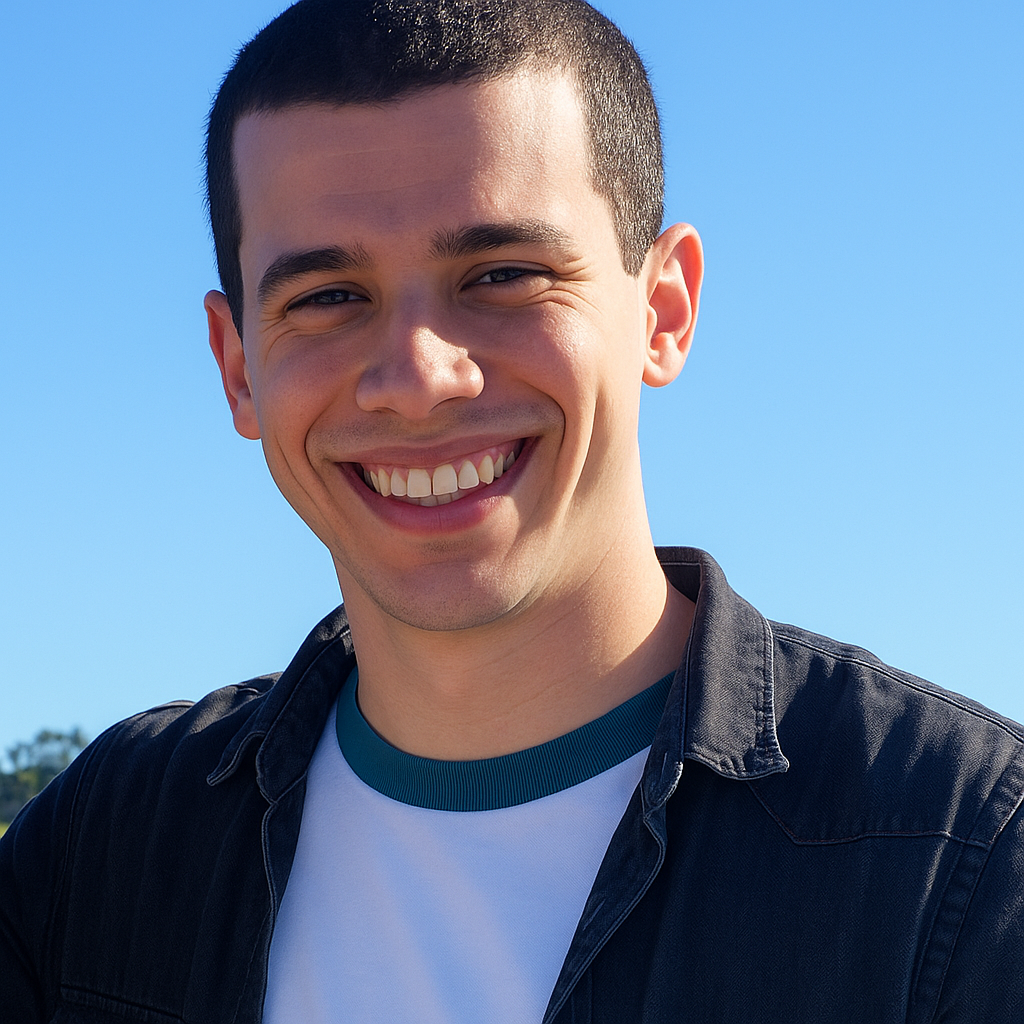}}]{Tiago da Silva Barros}
received the Msc. degree from Université Côte d'Azur in 2022. He is currently a Phd student in Université Côte d'Azur since 2022.
His interests are optimization problems for networking, such as scheduling and resources allocation and Machine Learning models.
\end{IEEEbiography}
\vspace{-2.0cm}
\begin{IEEEbiography}[{\includegraphics[width=1in,height=1.25in,clip,keepaspectratio]{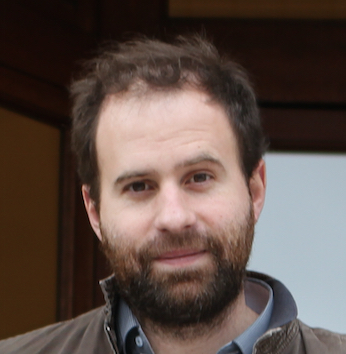}}]{Frederic Giroire}
currently is a senior research scientist at CNRS inside the joint team Coati between I3S (CNRS, University of Nice-Sophia Antipolis) laboratory and Inria which he joined in 2008. He received his Ph.D. from the University Paris 6 in 2006. He worked for 6 months in the research labs of Sprint (California) in 2002 and for one year in Intel Research labs (Berkeley) in 2007, leading to 3 patents. His research interests include algorithmic graph theory and combinatorial optimization for network design and management issues.
\end{IEEEbiography}
\vspace{-1.5cm}
\begin{IEEEbiography}[{\includegraphics[width=1in,height=1.25in,clip,keepaspectratio]{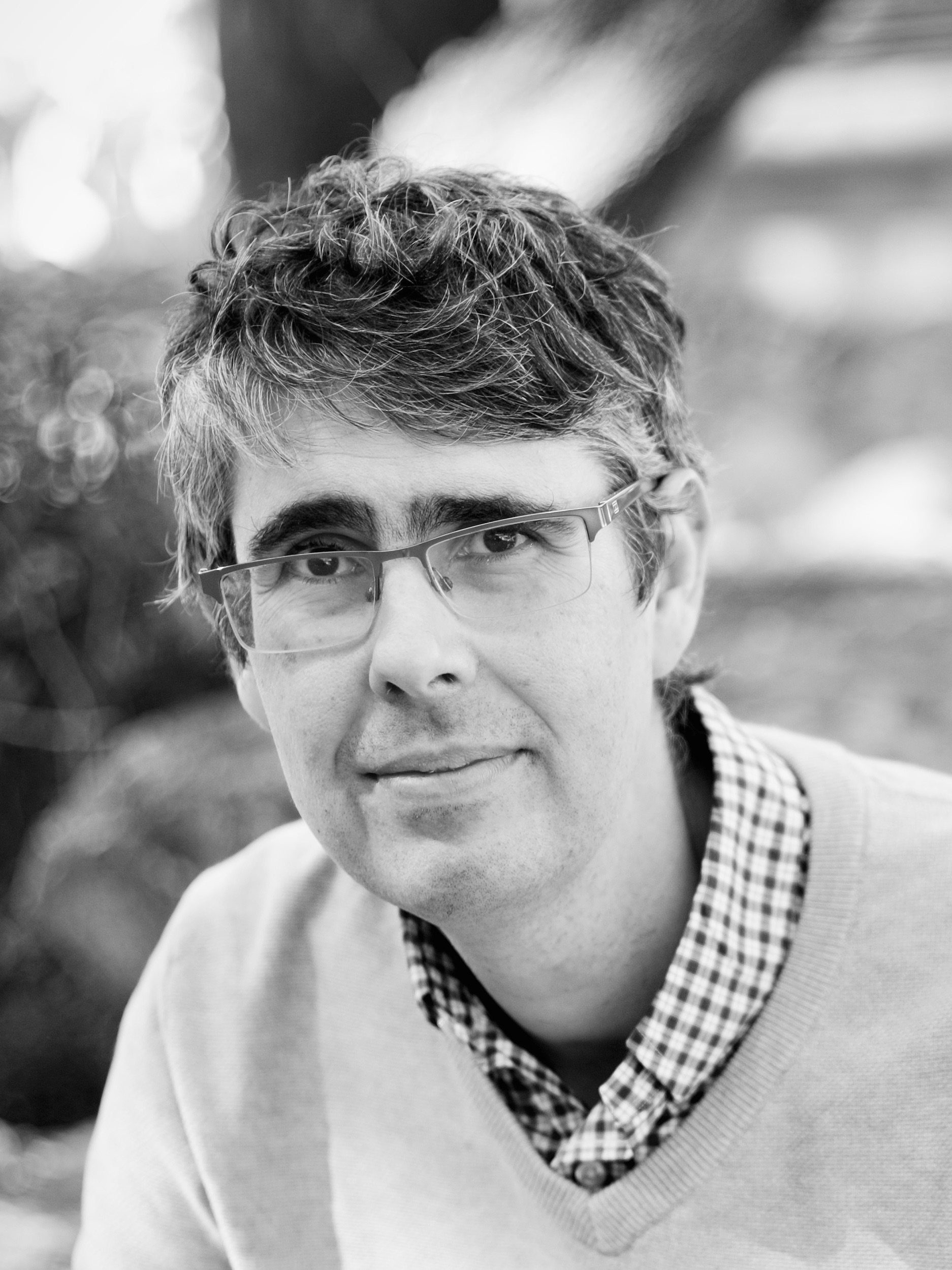}}]{Ramon Aparicio-Pardo}
received the MSc and
PhD degrees from Universidad Politécnica de
Cartagena (UPCT), Spain, in 2006 and 2011,
respectively. He is currently an associate professor with Université Côte d'Azur (UniCA), France,
since 2015. His PhD thesis was distinguished
with Telefonica Award for Best Thesis in Networking. His research interests include optimal design
and management of communication networks,
and more recently on machine learning-driven
network control.
\end{IEEEbiography}
\vspace{-1.5cm}
\begin{IEEEbiography}[{\includegraphics[width=1in,height=1.25in,clip,keepaspectratio]{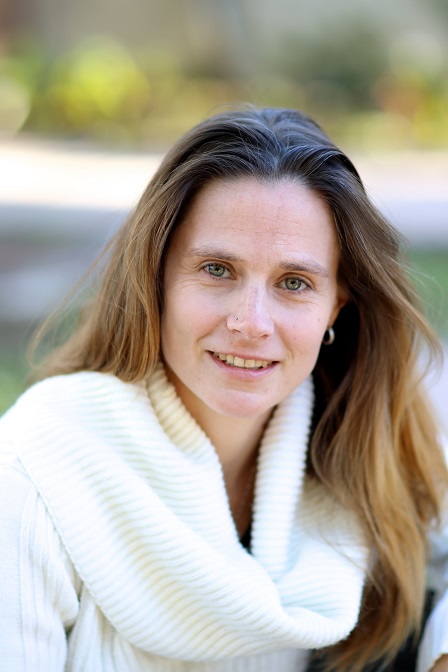}}]{Joanna Moulierac}
received the M.Sc. degree from the University of Montpellier, in 2003, and the Ph.D. degree in computer science from the University of Rennes, in 2006. She is  Associate Professor at Université Côte d'Azur (IUT Nice Côte d’Azur) since 2007. Her main research interests include algorithms, networks, combinatorial optimization, and energy-aware network designs and managements. 
\end{IEEEbiography}
\end{mainVersion}
\clearpage
\begin{appendixVersion}
\appendix

\section{Methods}

\subsection{CarbonTracker tool}
\label{sec:methods:carbonTracker}

\textit{CarbonTracker} \citeappendix{anthony2020carbontracker} is a software-based energy measuring tool designed for predicting the energy consumption and carbon footprint of training deep learning models.

It is implemented in python and it measures the GPU and CPU. For this, it utilizes the energy measuring interfaces \texttt{nvidia nvml} and \texttt{intel rapl}, respectively. Further information is described below.
\begin{tcolorbox}
    \nitbf{Nvidia Management Library (NVML) \citeappendix{nvidia2022nvidia}} is an interface which allows managing Nvidia graphical processing units (GPU). The interface allows measuring the power consumption of the GPU with the command \texttt{nvml Device Get Power Usage}, which returns the power usage for this GPU in milliwatts, with a $5\%$ of accuracy. 
\end{tcolorbox}
\begin{tcolorbox}
    \nitbf{\texttt{intel rapl} \citeappendix{khan2018rapl}}
    is an interface which allows users to estimate, monitor and manage the energy consumption in Intel processors.
Intel RAPL provides two main functionalities: (i) energy consumption measurement with a high sampling rate and (ii) limiting the maximal power consumption across the processing domains.
\end{tcolorbox}
The tool is widely used by researchers \citeappendix{geissler2024power, douwes2021energy, bannour2021evaluating, rodriguez2024evaluating}, is well documented, and provides a user-friendly interface. 

\subsection{Experimental Setup for Carbon Tracker validation}
\label{sec:methods:experimentalSetupCarbonTracker}

In this section, we describe the experimental setup for validating the \texttt{carbonTracker} tool.

\nitbf{Hardware configuration.} During our experiments, we used system equipped with an Intel Xeon Gold 6230R CPU and two NVIDIA RTX A5000 GPUs, running Ubuntu 24.04 as the operating system. Additionally, we used the power meter \emph{PowerSpy2} \citeappendix{powerspy2}, by Alciom. 

\nitbf{Tasks and models.} In our experiments, we measured the inference energy consumption by models across six tasks: \textit{object detection}, \textit{text generation}, \textit{speech recognition}, \textit{text classification}, \textit{text to image} and \textit{image-text to text}.


For each task, we selected models evaluated in the benchmarks available on the Hugging Face platform, as these models can be easily retrieved from there. For each task, we evaluated models of different sizes. However, for some tasks, we were unable to run the largest models on our hardware. Below, for each task, we list the benchmark, the number of models we ran, and the size (in terms of number of parameters) of the largest model we were able to run.
\begin{enumerate}
     \item \textbf{Text generation}: OpenLLM Leaderboard. 11 models. Largest model: 1.6B of parameters.  
    \item \textbf{Object detection}: Open Object Detection Leaderboard. 11 models. Largest model: 48M of parameters.
    \item \textbf{Speech recognition}: Open ASR Leaderboard. 17 models. Largest model: 1.5B of parameters.
    \item \textbf{Image-text to text}: MMMU. 4 models. Largest model: 4.2B of parameters.
    \item \textbf{Text to image}: GenAI. 3 models. Largest model: 3.4B of parameters.
    \item \textbf{Text Classification}: MTEB Leaderboard. 8 models. Largest model: 17M of parameters.
\end{enumerate}


In total, 54 models were evaluated. Each model was executed 10 times in order to obtain consistent results.

\smallskip


\nitbf{Input data.} In our experiments, we used the following input data:

\begin{enumerate}
    \item \textbf{Text generation}: An input prompt containing the following command: \textit{“Write a 2000-token story about a futuristic world where AI and humans coexist."}. 
    \item \textbf{Object detection}: A set of 100 images from Coco 2017 dataset.
    \item \textbf{Speech recognition}: A set containing 100 samples fom \textit{LibriSpeech} dataset.
    \item \textbf{Image-text to text}: Two samples (questions) from MMMU PRO dataset.
    \item \textbf{Text to image}: An input prompt with the command \textit{"Astronaut in a jungle, cold color palette, muted colors, detailed, 8k"}.
    \item \textbf{Text Classification}: A set containing 10000 samples from \texttt{imdb} dataset.
\end{enumerate}

\nitbf{Power meter measurement adjustment.}
Since the power meter measures total system power consumption, including fans, we took idle power consumption into account. We first measured the average power consumption of the system in an idle state (no models running) for 15 seconds. Then, during model execution, we subtracted this idle power from the total power measured by \textit{PowerSpy}, following the methodology of Rodriguez et al.~\citeappendix{rodriguez2024evaluating}.



\subsection{Experimental setup for energy consumption dependency on number of parameters}
\label{sec:methods:experimentalSetupNumberOfParams}

In this section, we describe the experimental setup for investigating the relation between the model size (number of parameters) and the energy consumption. \\

\nitbf{Hardware configuration} During our experiments, we used a system equipped with a Dell R7525 dual-AMD EPYC 7413 and Nvidia A40 GPU card. \\

\nitbf{Models} During our experiment, we performed inference requests over 241 models, according to the following division:
\begin{itemize}
    \item Image classification (vision): 213 models. The models were extracted from \textit{PyTorch Images (timm) benchmark}
    \item Speech recognition (audio): 17 models. Extracted from \textit{Open ASR leaderboard.}
    \item Text generation (language): 14 models. Extracted from \textit{llm-perf benchmark}. 
\end{itemize}

Each model was performed 10 times. \\

\nitbf{Input data.} In our experiments, we used the following datasets for performing the inference tasks.

\begin{itemize}
    \item Image classification (vision): A set of 100  images from ImageNet dataset.
    \item Speech recognition (audio): A set containing 50 audio samples from Librispeech dataset.
    \item Text generation (language): An input prompt containing the following command: \textit{"Write a 50-token story about a futuristic world where AI and humans coexist."}
\end{itemize}

\subsection{AI benchmarks methodology}
\label{Methods:benchmarksMethodology}

For this study, we selected relevant AI benchmarks that analyze and compare models addressing the investigated tasks. For this, we selected benchmarks which assess (i) the model size (in terms of number of parameters) and (ii) the utility value, which evaluates models performance. All the benchmarks 
are described in 
\begin{natureReferences}
    Methods \ref{Methods:benchmarksDescription}.
\end{natureReferences}
\begin{transactionsReferences}
    Appendix \ref{Methods:benchmarksDescription}.
\end{transactionsReferences}

\subsubsection{Benchmark sources}

In our investigations, we selected benchmarks from two different sources: \textit{Hugging Face} and \textit{Papers With Code} platforms.

\subsubsection{Number of parameters estimation}

Although compiling models size is one of our criteria for selecting benchmarks, in some benchmarks,  some models parameter count were not available.


For \textit{time series forecasting} task, the benchmarks using the\textit{ETTh1} dataset did not include parameter counts. Then, when available, we retrieved this information from the original research papers.

For the \textit{text generation} (more specifically \textit{Lmsys NPHardEval} benchmark), \textit{mathematical reasoning} and \textit{Image-text to text} tasks, some models do not publicly disclose their weights (e.g., \texttt{GPT-4} or \texttt{Gemini}). In such cases, we relied on size estimates provided by Ecologits \citeappendix{ecologits}.

\subsubsection{Models usage}

In our analysis, we wanted to measure model usage, i.e., the extent to which models are adopted by users in data centers. Since this information is not publicly available, we estimated model usage by looking at the number of downloads:
\begin{enumerate}
    \item For benchmarks hosted on the \textit{Hugging Face} platform, we considered the number of monthly downloads obtained through the \textit{Hugging Face} API; 
    \item For benchmarks from \textit{Papers With Code}, for each model, we searched for its implementation on the  \textit{Hugging Face} platform and, if available, we retrieved the number of monthly downloads; 
    \item For models that are not open-source and only accessible through web applications, we estimated popularity based on monthly website visits using data from \textit{SimilarWeb} \citeappendix{similarweb2025}.
\end{enumerate}
For the \textit{Papers With Code} \textit{translation} task benchmark, most of the papers, except for \texttt{T5-11B}, have no implementation in the \textit{Hugging Face} platform. Thus, we decided to extend the benchmark by adding other \texttt{T5} family models that were not considered in the original benchmark, as they are the most downloaded models for translation on \textit{Hugging Face}. We considered the utility values reported in the original paper \citeappendix{raffel2020exploring}, introducing the \texttt{T5} models.

\subsubsection{Key models selection}\label{Methods:Keymodelsselection}

For each evaluated task, we select among the available models in the corresponding benchmarks, the two key models (one best-performing and one energy-efficient). For this, we first define the efficiency metric as the ratio between the utility value and the number of parameters.
\[
\text{\textit{Efficiency}} = \frac{\text{\textit{Utility}}}{\text{\textit{Nb. Parameters}}}
\]

Then, we identify the key models using the following criteria:

\begin{enumerate}
    \item {\best} model: The model with the highest utility value.
    \item {\efficient} model: The model with the highest efficiency metric when the utility drop is below $5\%$. For tasks \textit{Text to Image}, \textit{Image-text to Text}, \textit{Speech Recognition}, and \textit{object detection}, there is no model satisfying the constraints, since discrete models are being evaluated. Then, in such cases, we consider the model with the highest efficiency, which has a utility drop of more than $5\%$, in the worst case reaching $18.2\%$.
\end{enumerate}

\subsection{Tasks and benchmarks description}
\label{Methods:benchmarksDescription}

To provide an overview of current trends and the categorization of tasks in Artificial Intelligence (AI), Table~\ref{tab:AIIndexFields} and Table~\ref{tab:listTasksFieldsHuggingFace} present the most commonly addressed AI fields and tasks reported in the AI Index Report~\citeappendix{perrault2024artificial} and on the \textit{Hugging Face} platform, respectively.

In Section II, we identify the most frequently addressed AI tasks in data center deployments. Below, we describe the specific tasks evaluated in our study, along with representative benchmarks commonly used to assess model performance in each task.

\begin{table*}[ht]
    \scriptsize
    \centering
     \begin{subtable}[b]{0.38\textwidth}
    \renewcommand{\arraystretch}{1.3} 
    \begin{tabular}{|>{\centering\arraybackslash}m{2cm}|>{\centering\arraybackslash}m{2cm}|}
         \hline
         \textbf{Field} & \textbf{Sub-field} \\
         \hline
         \multirow{3}{*}{\textit{Language}} & Understanding \\
         \cline{2-2}
         & Generation \\
         \cline{2-2}
         & Factuality\\
         \hline
         \textit{Coding} & Generation \\
         \hline
         \multirow{6}{*}{\shortstack{\textit{Image} \\ \textit{Computer}\\\textit{Vision}}} & Generation \\
         \cline{2-2}
         & Instruction Following \\
         \cline{2-2}
         & Editing \\
         \cline{2-2}
         & Segmentation \\
         \cline{2-2}
         & \shortstack{3D \\ Reconstruction} \\
         \hline
         \textit{Video Computer Vision} & Generation \\
         \hline
         \multirow{5}{*}{\textit{Reasoning}} & General \\
         \cline{2-2}
         & Mathematical \\
         \cline{2-2}
         & Visual \\
         \cline{2-2}
         & Moral \\
         \cline{2-2}
         & Casual \\
         \hline
         \textit{Audio} & Generation \\
         \hline
         \multirow{2}{*}{\textit{Agents}} & General \\
         \cline{2-2}
         & Task Specific \\
         \hline
         \textit{Robotics} & - \\
         \hline
         \textit{Reinforcement Learning} & - \\
         \hline
    \end{tabular}
    \caption{AI trending fields (with respective sub-fields) reported by The AI Index Report \cite{perrault2024artificial}}
    \label{tab:AIIndexFields}
    \end{subtable}
    \hfill
    \begin{subtable}[b]{0.58\textwidth}
    \centering
    \begin{tabular}{|c|c|}
    \hline
    \textbf{Task} & \textbf{Field} \\
     \hline
    Text Classification & Language\\
    Question Answering & Language \\
    Text Generation & Language \\
    Translation & Language \\
    Token Classification & Language \\
    Text2text Generation & Language \\
    Fill Mask & Language \\
    Sentence Similarity & Language \\
    Summarization & Language \\
    Multiple Choice & Language \\
    Zero Shot Classification & Language \\
    Text Retrieval & Language \\
    Text To Text & Language \\
    Feature Extraction & Language \\
    Table Question Answering & Language \\
    Table To Text & Language \\
    \hline
    Automatic Speech Recognition & Audio \\
    Audio Classification & Audio \\
    Text To Speech & Audio \\
    Audio To Audio & Audio \\
    Voice Activity Detection & Audio \\
    Text To Audio & Audio \\
    \hline
    Image Classification & Vision \\
    Object Detection & Vision \\
    Image Segmentation & Vision \\
    Zero Shot Image Classification & Vision \\
    Video Classification & Vision \\
    Image To Text & Vision \\
    Image Feature Extraction & Vision \\
    Image To Image & Vision \\
    Unconditional Image Generation & Vision \\
    Image Text To Text & Vision \\
    Image To Video & Vision \\
    Image To 3D & Vision \\
    Depth Estimation & Vision \\
    Visual Question Answering & Vision \\
    Document Question Answering & Vision \\
    Zero Shot Object Detection & Vision \\
    \hline
    Reinforcement Learning & Reinf. Learning \\
    \hline
    Tabular Classification & Tabular \\
    Tabular Regression & Tabular \\
    Time Series Forecasting & Tabular \\
    \hline
    Graph ML & Graph \\
    \hline
    Text To Video & Multimodal \\
    Mask Generation & Multimodal \\
    Text To 3D & Multimodal \\
    Video Text To Text & Multimodal \\
    Text To Image & Multimodal \\
    Image To Video & Multimodal \\
    Image Text To Text & Multimodal \\
    \hline
    Robotics & Robotics \\
    \hline
    Other & Other \\
    \hline
    \end{tabular}
    \caption{List of tasks and fields on the \emph{Hugging Face} platform.}
    \label{tab:listTasksFieldsHuggingFace}
    \end{subtable}
    \caption{\textbf{List of Artificial Intelligence (AI) fields and subfields or tasks.} (Left \ref{tab:AIIndexFields}) Reported by the AI Index Report and (Right \ref{tab:listTasksFieldsHuggingFace}) on the \textit{Hugging Face} platform.}
\end{table*}

\subsubsection{Text generation}

The text generation task consists of creating a text based on some input which contains the instructions. The text generated should be clear and coherent logically and grammatically.

\begin{tcolorbox}
\nitbf{Benchmark: \textit{Open LLM Leaderboard} \citeappendix{open-llm-leaderboard-v1}}
The benchmark was developed by \emph{Hugging Face}, using according to Eleuther framework \citeappendix{eval-harness}. The benchmarks evaluate the generative models in tasks such as the ability to follow specific formatting instructions, high-school mathematical problems, algorithmic generated problems and multiple-choice knowledge questions. 

\end{tcolorbox}

\begin{tcolorbox}
\nitbf{Benchmark: \textit{LMSys Chatbot arena Leaderboard }\citeappendix{chiang2024chatbot, zheng2024judging}}
This benchmark, developed by LMSYS and UC Berkeley, evaluates text generation models using \emph{chatbot arena}, using a score based on humans preferences.



    
\end{tcolorbox}

\subsubsection{Mathematical reasoning}

In mathematical reasoning, the systems are requested for solving a wide range of mathematical problems.

\begin{tcolorbox}
    \nitbf{Benchmark:\textit{ NPHard Eval Benchmark }\citeappendix{fan2023nphardeval} (mathematical reasoning)} The \emph{NPHard} benchmark evaluates the LLM power for reasoning in mathematical questions. It contains over than 900 questions and the objective is to classify the problems into 3 categories: P, NP-complete and NP-hard. 

\end{tcolorbox}


\subsubsection{Code generation}

In code generation tasks, the models, usually based on Natural Language Processing (NLP), are requested for generating source code. The tasks normally involve translating pseudocode to executable code, autocompleting functions and translating code between programming languages.

\begin{tcolorbox}
    \nitbf{Benchmark: \textit{BigCode Leaderboard \citeappendix{bigcode-evaluation-harness}} (code generation)} The \emph{bigCode} models leaderboard evaluates LLM for generating code scripts. The leaderboard evaluates over two main datasets: (i) \emph{HumanEval}, that analyzes the functional correctness of synthetizing python programs; and (ii) \emph{MultiPL-E}, which translates the previous dataset into other programming languages. 
\end{tcolorbox}

\subsubsection{Translation}

In text translation task, the models receive a task in a certain language, and they should generate the task translated into another language.




\begin{tcolorbox}
\nitbf{Benchmark: \textit{Machine Translation english-german} \citeappendix{bojar-etal-2014-findings}}
 The \emph{WMT (World Machine Translation)} is a dataset which evaluates pairs of languages. The main metric used is the \emph{BLEU score}, which measures how many \emph{n-grams} (sequences of words) in the machine-generated translation match those to a reference translation.



\end{tcolorbox}

\subsubsection{Text classification}

The text classification task consists of assigning a label or a category based on an input text. This task may be used for several applications such as spam detection, sentiment analysis, content classification and sentence similarity. Usually, the task involves data preprocessing, features extraction and then, the model training and evaluation.

\begin{tcolorbox}
    \nitbf{MTEB Leaderboard \citeappendix{muennighoff2022mteb}}
The \emph{MTEB} leaderboard evaluates LLM models for classification tasks. \emph{MTEB} evaluates over 58 datasets and 112 languages. This comprises 8 embedding tasks: Bitext mining, classification, clustering, pair classification, re-ranking, retrieval, STS and summarization.
\end{tcolorbox}

\subsubsection{Text Clustering}

In text clustering, the models are requested for grouping similar textual documents based on the content. The models aim to identify patterns and structures, forming clusters where samples with similar content are clustered in the same group. 

The benchmark used for this task was the \textit{MTEB Leaderboard} \citeappendix{muennighoff2022mteb}, as in \textit{text classification}.


\subsubsection{Object detection}

In the object detection task, the model must identify and locate multiple objects in an image.

\begin{tcolorbox}
\nitbf{Benchmark: \textit{Open Object Detection} \citeappendix{open-od-leaderboard}}
This benchmark was developed by \emph{HuggingFace}.
The dataset used is the \emph{Microsoft Common Objects in Context (COCO)}, containing 80 object categories, images with complex scenes, and high-quality and manually annotated labels.
The main metric used for evaluating the models is the average precision.




    
\end{tcolorbox}

\subsubsection{Image classification}

In the image classification task, the models should assign a label to an image based on its visual content. 
The main metric used is the accuracy, which counts the number of correct classified images over the total of images.

\begin{tcolorbox}
    \nitbf{Benchmark: \textit{ImageNet} \citeappendix{russakovsky2015imagenet}}
The \textit{ImageNet} dataset is a large-scale visual database for image classification. The subset used in this study, \textit{ImageNet-1k}, comprises over \SI{1.2}{M} images categorized into \num{1000} classes.



\end{tcolorbox}
We used two sources for the \textit{ImageNet} dataset as we tested both models reported in \textit{Hugging Face} and in \textit{Papers With Code} for this task: (i) from the \textit{PyTorch Timm Leaderboard} \citeappendix{rw2019timm}, which evaluates \texttt{Pytorch} models in the \textit{Hugging face} platform; and (ii) from the \textit{Papers With Code} platform addressing \textit{Image Classification} on \textit{ImageNet} \footnote{https://paperswithcode.com/sota/image-classification-on-imagenet}.

\subsubsection{Image segmentation}

In the semantic segmentation task, the goal is classifying each pixel in the image, providing a good understanding of the entire scene.

\begin{tcolorbox}
\nitbf{Benchmark: \textit{ADE20K} \citeappendix{Zhou_2017_CVPR}}
The \emph{ADE20K} consists of a dataset containing over 20k images and more 150 categories, such as cars, trees, and people.


    
\end{tcolorbox}

\subsubsection{Text to image}
In the text to image task, the models are requested for creating visual content from input text prompts. The goal is to produce high-quality images which can be used for several applications, such as art, design and medical imaging.

\begin{tcolorbox}
   \nitbf{Benchmark: \textit{GenAI} \citeappendix{ku2024imagenhub}} The \emph{GenAI} benchmark evaluates 17 models using ELO rating system based on public vote between the given results of 2 different models. The benchmark counts more than 8k votes for formulating the model score. 
\end{tcolorbox}


\subsubsection{Automatic speech recognition}

The speech recognition task consists of translating an audio extract into a text. It involves processing audio input, identifying the words spoken, and transcribing them accurately.

\begin{tcolorbox}
    \nitbf{Benchmark: \textit{Open ASR Leaderboard} \citeappendix{open-asr-leaderboard}}
The \emph{Open ASR Leaderboard} was developed by \emph{HuggingFace}.
For evaluating the models, the benchmark utilizes a set containing six datasets as described in \citeappendix{gandhi2022esbbenchmarkmultidomainendtoend}. 
The metric used is the \emph{Word Error Rate (WER)}, which computes the percentage of words in the system's output that are different from the reference transcript.

\end{tcolorbox}

\subsubsection{Audio classification}
The audio classification task involves assigning a label to an audio signal input. 

\begin{tcolorbox}
    \nitbf{Benchmark: \textit{ARCH Benchmark} \citeappendix{ARCH}}
\emph{ARCH} benchmark aims to benchmark models for audio classification task. The benchmark evaluates the models across 12 datasets with different input data: sound events (4 datasets), music samples (4 datasets) and speech samples (4 datasets).

\end{tcolorbox}









\subsubsection{Image-text to text}
In image-text to text tasks, the models are provided with images and text prompts as inputs and should output text. Also called as vision-langauge models (VLMs), the models used in this tasks are widely used for several applications, such as multimodal dialogue, visual question answering and image recognition.


\begin{tcolorbox}
    \nitbf{Benchmark: \textit{MMMU benchmark} \citeappendix{yue2023mmmu}}
The MMMU benchmark evaluates multimodal models on massive multidiscipline tasks demanding college-level subject knowledge and reasoning. \emph{MMMU} includes 11.5K multimodal questions from college exams.

\end{tcolorbox}


\subsubsection{Time series forecasting}

The time series forecasting task consists of a regression task, in which the AI models should predict the future elements of a time series based on historical trends in the past.

\begin{tcolorbox}
    \nitbf{Benchmark:\textit{ etth1-336} \citeappendix{gemmeke2017audio}}
The \emph{Etth1} benchmark evaluates the models on forecasting a time series. The time series records the operation and environmental characteristics of electricity transformers along time. Each data point consists of the target value ``oil temperature” and six power load features.
\end{tcolorbox}


\subsection{Experimental setup for estimating energy consumption}
\label{sec:methods:energyEstimationExperimentalSetup}


To estimate the inference energy consumption of a model when direct measurement was not feasible, we employed a regression-based approach based on empirical energy measurements. Our methodology involved evaluating the two key models across several AI tasks.

\smallskip

\nitbf{Hardware Configuration.}
Experiments were conducted on a Dell R7525 system equipped with a dual-AMD EPYC 7413 processor and an Nvidia A40 GPU.

\smallskip

\nitbf{Tasks and input data.}
We considered the following tasks in our experiments:
\textit{text generation, image classification, object detection, speech recognition, text-to-image generation, image-text-to-text generation, text classification, translation, image segmentation, audio classification,} and \textit{time series forecasting}.

Some task labels, specifically \textit{text clustering}, \textit{mathematical reasoning}, and \textit{code generation}, are not explicitly available in \textit{Hugging Face}. Therefore, we approximated these as follows: \textit{text clustering} was mapped to \textit{text classification}, while \textit{mathematical reasoning} and \textit{code generation} were treated as \textit{text generation} tasks.

\noindent For each task, we used the following datasets to run the inference requests. The datasets come from the ones used in the considered \textit{Hugging Face} and \textit{Paper with codes} benchmarks, except for \textit{Text Generation}, which uses a code example for one of the models.

\begin{enumerate} 
    \item \textbf{Image classification}: A set of 500 images of CIFAR-10 dataset.
    \item \textbf{Object detection}: A set of 100 images from Coco 2017 dataset.
    \item \textbf{Speech recognition}: A set containing 100 samples fom \textit{LibriSpeech} dataset.
    \item \textbf{Image-text to text}: A single sample (question) from MMMU PRO dataset.
    \item \textbf{Text to image}: An input prompt with the command \textit{"Astronaut in a jungle, cold color palette, muted colors, detailed, 8k"}.
    \item \textbf{Translation}: An input prompt command for translating from english to german the following sentence: \textit{This is a beautiful word.}
    \item \textbf{Image segmentation}: A sample containing one image from \textit{Coco 2017 dataset}.
    \item \textbf{Audio classification}: A set containing 10 samples from \textit{LibriSpeech} dataset.
    \item \textbf{Time series forecasting}: A set containing 17420 entries from \textit{etth1} dataset.
    \item \textbf{Text Classification}: A set containing 10000 samples from \texttt{imdb} dataset.
    \item \textbf{Text generation}: An input prompt containing the following command: \textit{“Write a 50-token story about a futuristic world where AI and humans coexist."}.
\end{enumerate}

\nitbf{Energy Measurement and Regression Analysis.}
For each task, we used the \textit{CarbonTracker} tool to measure the inference energy consumption of the two key models: the \textit{energy-efficient} model and the \textit{best-performing} model. Each model was evaluated over five runs.

Using the collected energy consumption data, we performed a linear regression on a log-log scale, following the pattern described in Section III-A.
The regression model follows:
\[
\log_{10}(E) = \alpha \cdot \log_{10}(P) + \beta, 
\]
where $E$  represents the inference energy consumption and $P$ denotes the number of model parameters. The coefficients $\alpha$ and $\beta$ were estimated for each task ($\alpha$ ranges from \num{0.27} to \num{0.84} with a median value of \num{0.59}, when $\beta$ ranges from \num{-5.89} to \num{1.4} with a median value of \num{-2.56}).

With these coefficients, the energy consumption of other models within the same task can be estimated using:
\[
E = 10^\beta \cdot P^\alpha. 
\]
The formula enables the estimation of the energy consumed by models other than those directly measured. 

\nitbf{Data availability}
All the benchmarks and models evaluated are publicly available on \textit{Hugging Face} and \textit{Papers with Code} platforms. We provide the data compiling all models information for every task on 
\hyperlink{https://github.com/tsb4/small-is-sufficient}{https://github.com/tsb4/small-is-sufficient}. 


\smallskip

\nitbf{Code availability.}
We provide the code used for running inference requests for each of the evaluated tasks. The code retrieves a dataset, load a model and performs the inference task measuring the energy consumption using \textit{CarbonTracker} tool. Our code is available at 
\hyperlink{https://github.com/tsb4/small-is-sufficient}{https://github.com/tsb4/small-is-sufficient}.

\clearpage
\bibliographystyleappendix{IEEEtran}
\clearpage   
\clearpage   

\end{appendixVersion}

\end{document}